\documentclass[12pt,preprint]{aastex}

\begin{document}
\title{Neutrinos from Fallback onto Newly Formed Neutron Stars}

\author{Chris L. Fryer\altaffilmark{1,2}}

\altaffiltext{1}{Department of Physics, The University of Arizona,
Tucson, AZ 85721} 
\altaffiltext{2}{CCS Division, Los Alamos National Laboratory, 
Los Alamos, NM 87545}

\begin{abstract}

In the standard supernova picture, type Ib/c and type II supernovae
are powered by the potential energy released in the collapse of the
core of a massive star.  In studying supernovae, we primarily focus on
the ejecta that makes it beyond the potential well of the collapsed
core.  But, as we shall show in this paper, in most supernova
explosions, a tenth of a solar mass or more of the ejecta is
decelerated enough that it does not escape the potential well of that
compact object.  This material falls back onto the proto-neutron star
within the first 10-15 seconds after the launch of the explosion,
releasing more than $10^{52}$\,erg of additional potential energy.
Most of this energy is emitted in the form of neutrinos and we must
understand this fallback neutrino emission if we are to use neutrino
observations to study the behavior of matter at high densities.  Here
we present both a 1-dimensional study of fallback using
energy-injected, supernova explosions and a first study of neutrino
emission from fallback using a suite of 2-dimensional simulations.

\end{abstract}


\section{Introduction}

The collapse of a massive star down to a neutron star releases over
$10^{53}$\,erg of potential energy.  Just one percent of this energy
is required to power the observed type Ib/c and type II supernovae.
Most of the energy is emitted in the form of neutrinos.  How this
small fraction of the energy is converted into explosion energy is
still a matter of debate, but it is believed by most that neutrinos
play a role in depositing energy above the collapsed core to drive
this explosion (see Fryer 2003 for a review).  Whether or not
neutrinos are important for the explosion, they do provide astronomers
a window into the explosion mechanism behind supernovae.

Neutrinos also provide a window into the material conditions at the
core of a supernova explosion.  With temperatures above 10\,MeV and
densities above nuclear densities, core-collapse supernovae make ideal
laboratories for nuclear physics.  To extract information about
nuclear physics from such laboratories, we must understand the
systematics in our supernova experiment.  When the explosion engine is
active, convective motions in the engine (Herant et al. 1994) make it
very difficult to interpret the neutrino signal.  An observed elevated
neutrino luminosity could be a change in the neutrino opacity or it
could be a difference in the convective engine.  One approach to
eliminate issues with convection would be to wait until all convective
activity ceases.  A number of studies have now been presented
following the evolution of a collapsing star over 0.5-1.5\,s after
bounce (Fryer \& Heger 2000; Burrows et al. 2006; Scheck et al. 2006;
and, in 3-dimensions, Fryer \& Young 2007).  In all these
calculations, hydrodynamic motions near the proto-neutron star
continue to dramatically affect the neutrino emission through the end
of the simulations.  To achieve a clean neutrino signal, we must wait
until after the launch of the explosion.  We must also wait a few
seconds after the launch of the explosion because the proto-neutron
star may experience deep convection (Keil et al. 1996).  After this
time (roughly a few seconds to 10-20\,s), the supernova finally seems
to have achieved an ideal condition as a physics laboratory (Reddy et
al. 1999).  After 10-20\,s, the neutrino signal will be too weak to do
much experimental science, so we truly are limited to this narrow time
window.

Unfortunately, even at these late times, Nature does not allow
completely pristine conditions.  Material falling back from the
supernova explosion (``fallback'') may well produce a new round of
convection and confusion to our neutrino signal.  This fallback has
been studied both in its important role in calculating the initial
mass of the neutron star formed in a supernova explosion (e.g. Fryer
\& Kalogera 2001) and in estimating the r-rpocess yields in supernovae
Fryer et al. (2006). In this paper, we study its role in determining
the neutrino luminosity arising after the first few seconds of a
supernova.  In \S 2, we review the history of supernova fallback and
present new calculations of supernova explosions estimating the
fallback for a range of explosion energies and stellar masses.  \S 3
describes the 2-dimensional code used to model the neutrino emission
from the this fallback and shows results for the suite of simulations
run for this paper.  Different than the multi-dimensional simulations
modeling stellar collapse, these simulations do not start at the onset
of collapse (or bounce) and end at the launch of the explosion.
Instead, they start a 2-10\,s (depending on the explosion energy,
etc. from \S 2) after the launch of the explosion when material begins
to fall back onto the neutron star.  We conclude with a brief
discussion on how neutrino signals can be used to help better
understand the supernova explosion, neutron star birth masses and
neutrino cross-sections.

\section{Supernova Fallback}

The idea of fallback was first discussed by Colgate (1971) to overcome
nucleosynthesis issues arising from the supernova ejection of neutron
rich material produced in stellar cores (Arnett 1971, Young et
al. 2006).  Colgate argued that the inner layers of the ejected
material would deposit its energy to the stellar material above it,
ultimately reducing its energy below that needed to escape the neutron
star, and it would fall back onto the neutron star.  In such a
scenario, one would expect the inner material to fall back quickly
(within the first few to ten seconds).  It was argued that this
material (the neutron rich material from the initial explosion) would
accrete onto the neutron star, alleviating any nucleosynthesis issues.

Since this work by Colgate, supernova explosion calculations have
confirmed that fallback does occur (Bisnovatyi-Kogan \& Lamzin 1984;
Woosley 1989; Fryer et al. 1999; MacFadyen et al. 2001) and new
arguments for the cause of this fallback were suggested.  For example,
Woosley (1989) argued that when the supernova shock decelerates in the
hydrogen layers of the star, it sends a reverse shock that drives
fallback.  Such a model argues that the fallback will happen at late
times, long after it can affect the neutrino luminosity from the
cooling proto-neutron star.  It also suggested that close binary
systems (where a star's hydrogen envelope was removed prior its
collapse) might experience a very different amount of fallback than
the amount of fallback in the collapse of a single star.

Which is the true cause of the fallback?  And more to the point for
neutrino emission, when does fallback occur?  This confusion has
mostly exists because, without a quantitatively reliable
explosion mechanism, scientists have artificially driven supernova
explosions to be able to study the results of these explosions
(e.g. supernova light curves, nucleosynthetic yields, and fallback).
Much of the past work (e.g. Woosley 1989; Fryer et al. 1999; MacFadyen
et al. 2001) used piston driven explosions.  By moving the piston out,
scientists artificially lowered the amount of fallback and delayed
this fallback considerably.  These calculations all predicted that
fallback would occur more than 100\,s after the launch of the
explosion.

But the piston-driven explosion mechanism may not accurately model the
nature of the fallback.  Young \& Fryer (2007) found that
piston-driven explosions produced both different fallback rates and
different nucleosynthetic yields than energy-driven explosions of the
same final energy.  The errors in the yields or light-curves are on
par with slight changes in the explosion energy.  Such small errors
seemed unimportant in matching the observations.  For fallback, the
differences are much more dramatic.  Not only does the amount of
fallback change, but the timescale at which the fallback occurs can
change by more than an order of magnitude, moving a falback time of a
few hundred seconds down to just 3-15\,s.  This changes the fallback
accretion rate by over an order of magnitude and ultimately determines
whether or not fallback is important in estimating the observed
neutrino signal.  It also means that the the amount of fallback will 
not change between binary and single stars (unless binary interactions 
change the internal structure of the star).

In this paper, we focus on fallback calculations from energy-injected
explosions.  Energy-injection is much better at mimicking the
currently-favored supernova mechnanisms.  Let's take the
convection-enhanced, neutrino-driven supernova engine as an example
(see Fryer 2003 for a review).  In this mechanism, the basic energy
source is neutrinos (either diffusing out of the proto-neutron star
core or from newly accreting material) that heat the atmosphere above the
neutron star.  This atmosphere is topped by the accretion shock of the
infalling star.  When enough energy is deposited in this region, the
accretion shock will be pushed outward and an explosion occurs.  
Convection aids this mechanism by both allowing heated material to
rise (cooling by adiabatic expansion instead of forcing it to continue
to heat until it can cool by neutrino emission) and allowing shocked
material at the top of the convective region to flow down to the
neutron star surface to accrete onto the neutron star (emitting
neutrinos) or be heating to be part of the rising bubble.  

An energy injection method, although not mimicking the effects of
convection directly, can mimick the basic tenets of this model -
heating just above the neutron star surface to blow off the accretion
shock.  Ideally, once the region above the neutron star becomes more
rarefied, the energy injection will essentially halt (aside from a
weak, by supernova standards, neutrino-driven wind).  Some groups have
gone so far as to only inject this energy through a neutrino flux
(Fr\"olich et al. 2006).  In this manner, once the region becomes
rarefied, the energy injection drops naturally.  A piston-driven
explosion can not mimic this effect well unless very specific
attention is paid to the input of the piston.  The very different
fallback results from piston explosions demonstrate just how far off
such explosions are from the energy drive of the standard neutrino
model.

This does not mean that by using energy-injection, we can produce a
definitive fallback estimate for a supernova explosion.  The explosion
energy is one of our primary uncertainties in estimating the fallback
rate.  Supernova scientists have yet to agree on the exact mechanism
behind the supernovae and we are far from achieving quantitative
predictions of these explosions.  With accurate light-curve and
spectra calculations, we may be able to estimate the explosion energy
for a particular supernova based on its observations.  By studying 
a range of explosion energies for a given progenitor, we can provide 
a template for the neutrino flux (from fallback) arising from 
these systems, allowing us to possibly extract this effect and 
once more focus a study on the physics of dense, hot matter.

\subsection{Calculating Fallback}

Our fallback calculations will all be based on fallback estimates from
energy-driven explosions.  We use the same multi-step technique
employed in Young et al. (2006) and Young \& Fryer (2007).  We start
with progenitor stars modeled to collapse (either from Heger et al.
2000 or Young et al. 2008).  These progenitors are mapped into the
1-dimensional core-collapse code from Herant et al. (1994).  This code
includes equations of state valid from stellar densities up to nuclear
densities, a 3-flavor flux-limited diffusion neutrino transport
scheme, and a simple nuclear network.  With this code, we follow the
collapse and formation of the proto-neutron star and the propogation
of the shock produced when the collapse halts due to nuclear forces.

After this shock stalls (as it loses its energy via neutrino losses),
we have a structure defined by a shocked ``atmosphere'' produced by
the now-stalled bounce shock above a dense proto-neutron star.  The
proto-neutron star typically has a baryonic mass between 1.1 and
1.3M$_\odot$.  The edge of the proto-neutron star is determined by the
mass where the density drops below $10^{10} {\rm g cm^{-3}}$.  At
these densities, there is typically a well-defined edge where the
density drops from $10^{12} {\rm g cm^{-3}}$ to $10^8 {\rm g cm^{-3}}$
over a very narrow mass cut.  The exact location of this edge depends
upon the progenitor mass and probably the exact code used to model
this collapse phase (e.g. 1-dimensional versus multi-dimensional
results).

In general, 1-dimensional calculations have not produced supernova
explosions.  With the proto-neutron star removed, we have also removed
the energy source for any explosion.  To induce an explosion in
1-dimension, we must source in energy.  The simulations here source
the energy directly into the inner 15 cells (roughly 0.1M$_\odot$) of
the star.  We keep this energy source on for a limited time (between
50 and 300\,ms), varying the energy injection rate and time to produce
a range of explosion energies.  Young \& Fryer (2007) found that such
energy sourcing was more flexible than a simple neutrino enhancement
to modeling the full range of proposed explosion mechanisms.  In our 
calculations, we use the shorter (50\,ms) injection duration for 
the lowest energy explosions and the lowest mass stars and the longer 
(300\,ms) injection duration for the more energetic explosions.

We follow this explosion for 400\,s.  This allows us to follow the
shock as it moves well into the star (and, for binary systems, out of
the star surface).  Our simulation space includes enough matter to
ensure that for this 400\,s duration, the shock is well within the
simulation space (there are at least 100 zones between the final shock
postion and the outer zone of our star).  In the case of the binary
systems, we have included a mass loss estimated from the stellar
models.  Typically, we model this star with roughly 2000 zones.  Young
et al. (2008) have done a resolution study, comparing 1000,2000 and
4000 zones.  They do not find an appreciable difference in the ejected
(and hence fallback) mass based on this resolution.  It will make a
large difference on how the fallback mass tapers off.  However, note
that with the Lagrangian code in this paper, we can not accurately
calculate low fallback rates.  But, as we shall see in our
2-dimensional models, it may well be that fallback powered explosions
might cut off this low-rate fallback.

We use the same 1-dimensional code to model the fallback.  As material
falls back onto the edge of our proto-neutron star and its density
rises above $10^9 {\rm g cm^{-3}}$, we remove the particle and add its
mass to our proto-neutron star.  Although we could follow the
accretion and neutrino cooling of this material to higher densities
with our code, as the density rises, the sound speed increases and the
cell size decreases, both of which cause the time step to decrease.
To make this problem tractable, we choose a low enough density that
the material accretes before dropping the timestep below 1
microsecond.  Even so, these simulations typically take between 1
million to 10 million timesteps.  As we shall find in \S 3, the 
modeling the true behavior of this matter requires modeling the 
accretion in multi-dimensions.  Since these simulations are 
focused on calculating the infall rate (not true accretion rate), 
our assumptions do not introduce large errors in our analysis.

In this paper, we present the results from a suite of 1-dimensional
explosion models models using 3 different progenitor masses and
explosion energies (for a summary, see Table~\ref{tab:fallback}).
Table~\ref{tab:fallback} also shows the peak accretion rates and total
mass accreted for these models.  Note that we predict a range of
neutron star masses based on both progenitor mass and explosion
energy, in agreement with Fryer \& Kalogera (2001).  But remember that
the neutron star masses assume that all of the fallback remains on the
neutron star.  As we shall see in \S 3, some of this matter is
re-ejected.  For normal explosion energies, it is likely that
12-15\,M$_\odot$ stars produce neutron stars with gravitational masses 
in the 1.3-1.5\,M$_\odot$ range.  If the explosions are stronger, the 
gravitational masses may well be as low as 1.2\,M$_\odot$.  For 
weak explosions, or more-massive stars, the remnant masses may 
be so large that the neutron star collapses to a black hole.

\begin{figure}
\plottwo{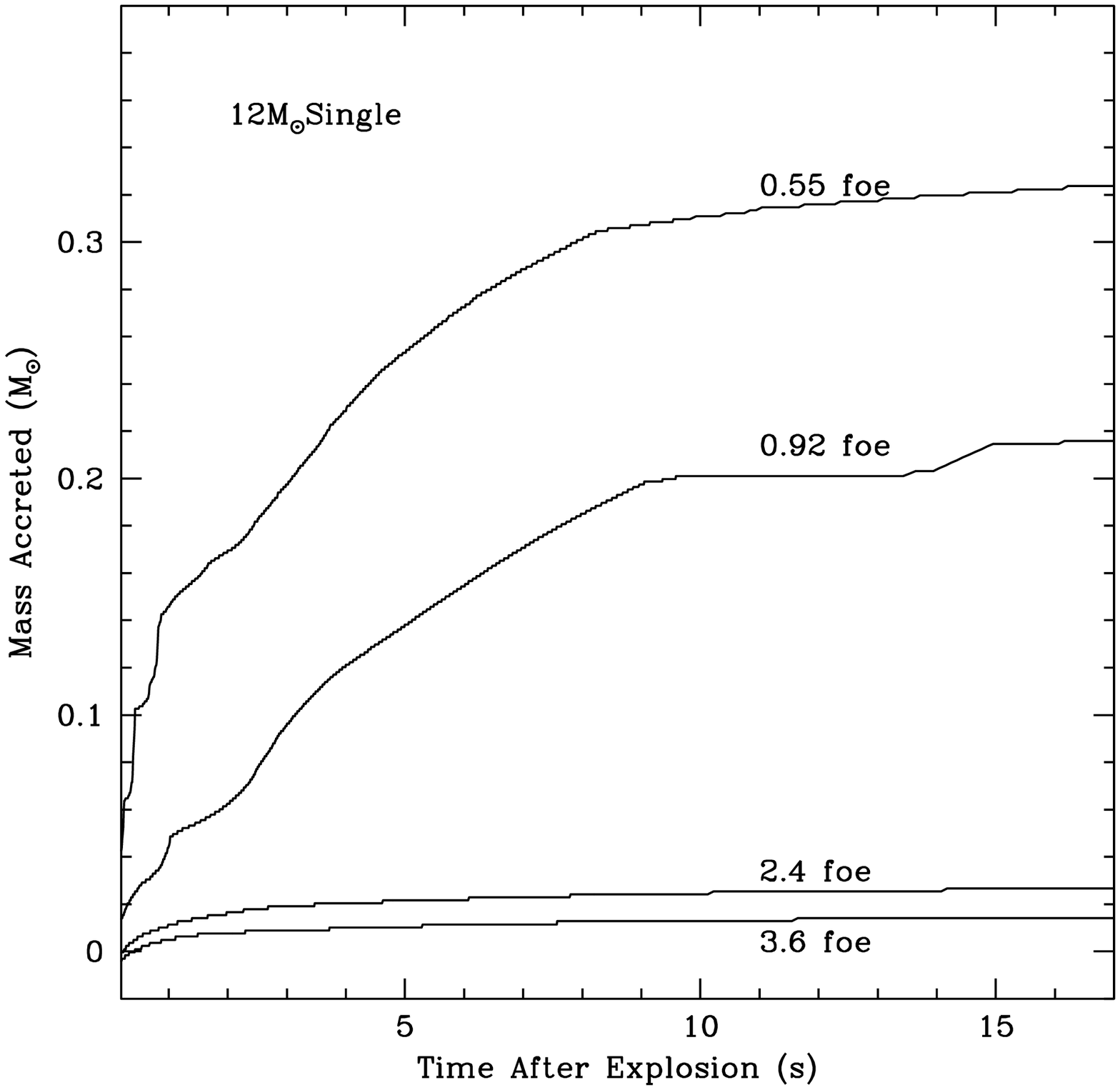}{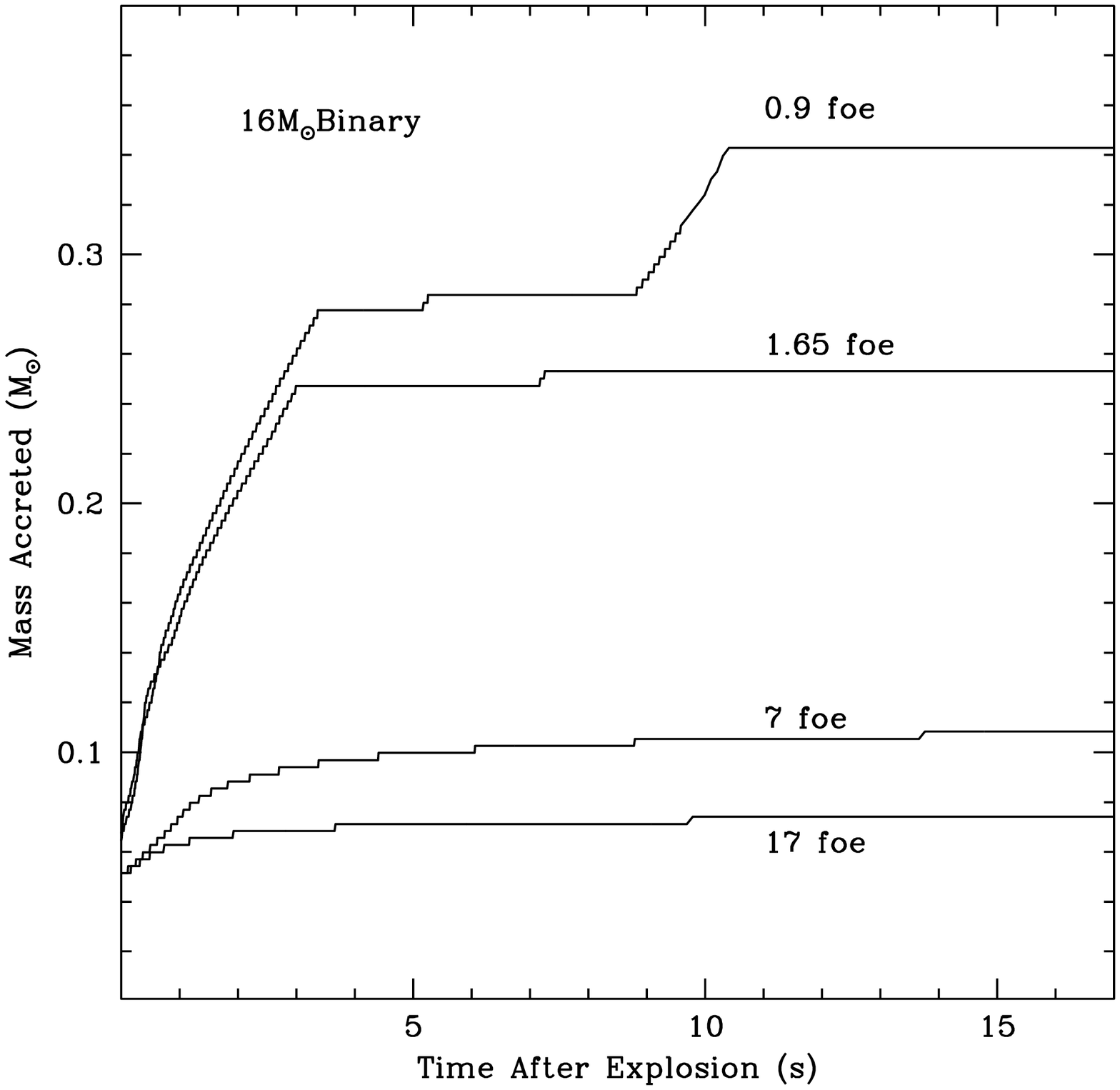}
\epsscale{0.5}\plotone{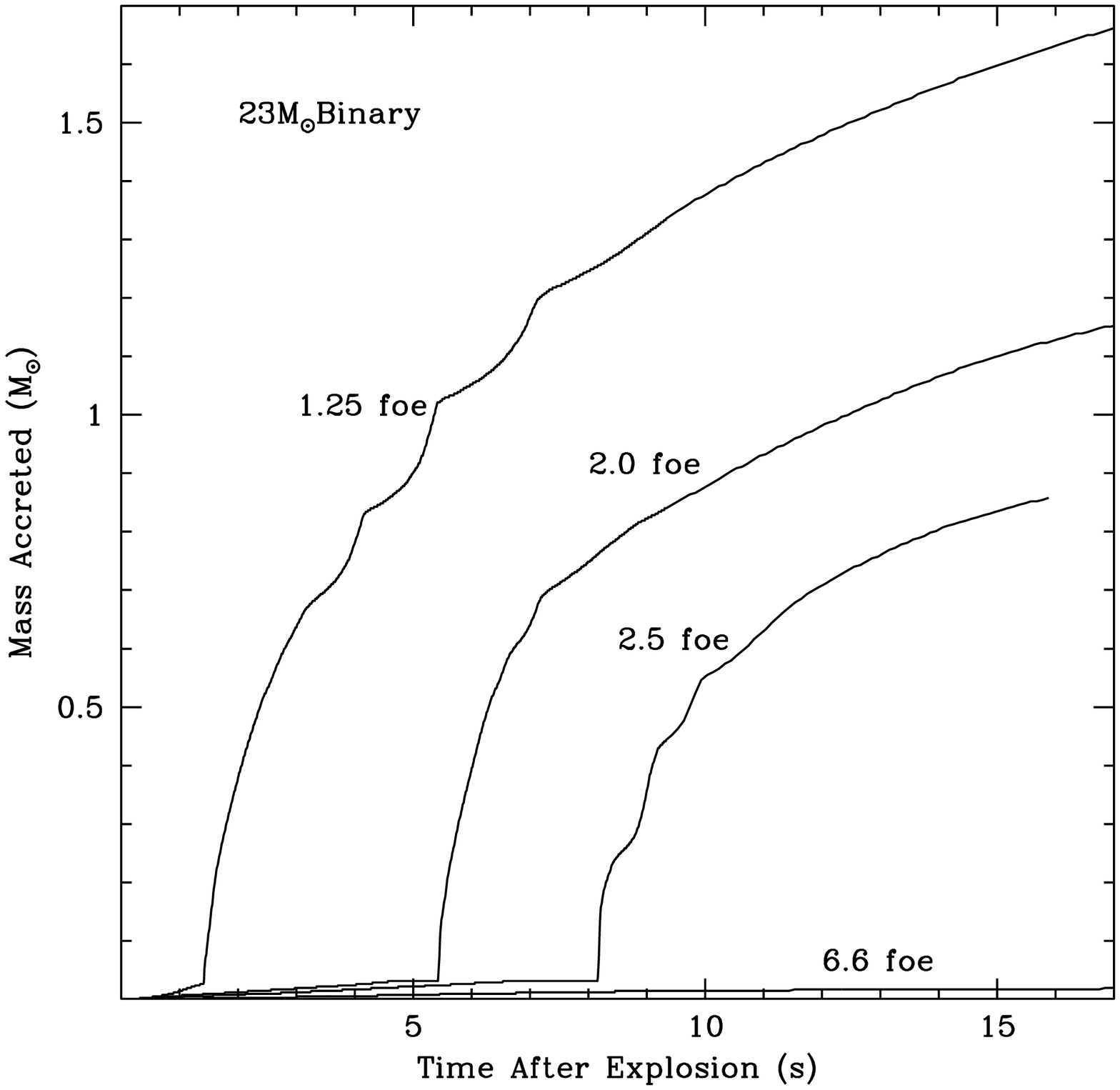}
\epsscale{1.0}
\caption{Fallback mass as a function of time from the launch of the
explosion for 3 different progenitor masses (one progenitor for each
panel) and a range of explosion energies (1 foe = $10^{51} {\rm
erg}$).  Most of the fallback is finished after 15\,s.  Note that for
a given explosion energy, the more massive the progenitor, the more
fallback that occurs.  But also note that for a standard ``1 foe''
explosion, our 12\,M$_\odot$ model still accretes 0.2\,M$_\odot$ in
roughly 10\,s corresponding to an accretion rate of 0.02\,M$_\odot
{\rm s^{-1}}$.}
\label{fig:macc}
\end{figure}

Figure~\ref{fig:macc} shows the mass accreted in the first 15\,s for
our models.  Note that in all casses, fallback occurs almost
immediately (with some delays of 2-7 seconds).  Unless large amounts
of fallback occur (above 1\,M$_\odot$), the fallback is likely to be
mostly over after 10-15\,s.  With over 0.1\,M$_\odot$ falling back in
10-15s, accretion rates above 0.01M$_\odot$\,s$^{-1}$ are expected,
corresponding to neutrino luminosities in excess of $10^{51} {\rm ergs
\, s^{-1}}$.  The bottom line is that, for normal supernovae, fallback
occurs, at least for stars with initial masses at or above
12\,M$_\odot$.  This fallback occurs early, so it will definitely play
some role in the neutrino emission at the 3-15\,s timescale.

\subsection{Angular Momentum}

An additional feature of the progenitor that affects the fallback and
the neutrino luminosity is the angular momentum in the progenitor.
Figure~\ref{fig:1drot} shows the angular momentum of the inner
4\,M$_\odot$ of a star for a variety of stars with and without
magnetic braking (Heger et al. 2000,2005).  For neutron star
accretion, the relevant angular momentum is that within the inner
1.4-2.0\,M$_\odot$.  Such mass zones have low-angular momenta: a
few$\times 10^{15} {\rm cm^2 s^{-1}}$ for stars with magnetic braking,
a few$\times 10^{16} {\rm cm^2 s^{-1}}$ for stars without.  Our
typical simulations use a value of $10^{16} {\rm cm^2 s^{-1}}$.  These
low angular momenta had little effect on our results, and we ran some
simulations with twice that amount.  For our black hole systems, we
ran even higher angular momenta.  For neutron stars, the angular
momentum is not enough for this material to form a true disk in the
star, but it can alter the downflow and it is this effect that we
would like to study in this paper.

Note that although there is quite a bit of structure in the angular
momenta (caused by incomplete angular momentum transport across
elemental boundaries), in general, the angular momenta of the stars
increases as one moves to higher and higher mass shell.  This is one
reason why black-hole forming systems are more likely to produce
asymmetric explosions than typical neutron-star forming systems.  The
increased angular momentum means that angular momentum plays a bigger
role in shaping the explosion, possibly producing larger asymmetries.

\section{Neutrinos from Fallback\label{results}}

Our 1-dimensional calculations provide us with a fallback rate.  If we
assume that any fallback material emits all of the potential energy
released from its downfall at the moment it hits the proto-neutron
star, we can estimate the neutrino luminosity from the fallback.  The
accretion of 0.1\,M$_\odot$ masses onto a 10\,km, 1.4\,M$_\odot$
neutron star over 10\,s would correspond to a neutrino luminosity of
$3.7\times 10^{51} {\rm ergs \, s^{-1}}$.  For our more massive stars,
this fallback can be ten times higher, corresponding to a neutrino
luminosity of $4\times 10^{52} {\rm ergs \, s^{-1}}$ for over 10\,s
after the launch of the explosion.  Assuming pair-annihilation
dominates the neutrino emission, this neutrino luminosity would be
nearly 50\% electron and 50\% anti-electron neutrinos.  Typical
neutron star luminosities after 1\,s are below $10^{52} {\rm ergs \,
s^{-1}}$ and can be as low as a few $\times 10^{51} {\rm ergs \,
s^{-1}}$ (e.g. Bruenn 1987, Keil \& Janka 1996).  Even at 1\,s,
fallback can dominate the neutrino emission if the fallback is heavy.
After $\sim$5\,s our fallback estimates argue that fallback neutrinos
dominate the neutrino emission.

But we have made several assumptions in this estimate for the neutrino
luminosity.  First, we have assumed that all of the potential energy
released is immediately emitted in neutrinos.  This energy can go into
heating the proto-neutron star which will then cool on a longer
timescale.  This energy also may go into ejecting other infalling
material.  As the material falls onto the proto-neutron star, it is
shock heated and can rise.  Fryer et al. (1996,2006) found that these
rising shocked bubbles can accelerate above the escape velocity and
actually be re-ejected.  So it is quite possible that a fraction of
the ``fallback'' material does not end up on the neutron star if we
account for these multi-dimensional effects.  This re-ejected material
both does not contribute to the total energy available to produce
neutrinos, but it takes some fraction of the potential energy released
to power its explosion.  Finally, $\mu$ and $\tau$ neutrinos may also 
be released and we must include these neutrinos in our energy budget.
In this section, we study the effects of the above assumptions to get
a more accurate neutrino luminosity from fallback.  

\subsection{Code Description}

Our primary concern with our simple estimate of fallback neutrino
emission is the fact that we ignore multi-dimensional effects.  We
have known for some time that if a compact object is accreting mass
with considerable angular momentum (and inefficient cooling), outflows
occur (e.g. Blandford \& Begelman 1999).  But even if the angular
momentum is minimal, if the compact object is a neutron star, a
sizable fraction of the infalling material can be re-ejected (Fryer et
al. 2006).  To study the fate of supernova fallback, we want to
understand a number of effects.  For example, how do the results vary
with accretion rate?  But we would like to also understand the role of
angular momentum and the differences between hot or cold neutron
stars.  Finally, numerical effects, such as boundary conditions, are
bound to play a role.  Before we discuss the results of our
simulations, let's discuss the code used for these calculations and
our tests of the physical and numerical effects.

Our code must model the physics of downflows and allow us to answer
the questions in the preceding paragraph.  First and foremost, we must
follow the evolution in a multi-dimensional manner.  As a first step,
we use the two-dimensional smooth particle hydrodynamics code
described in Fryer et al. (1996,2006).  We model the region from
10,000\,km above the proto-neutron star down to the proto-neutron star
surface.  Typical runs range from an initial set of $\sim 11,000$
particles moving up to 50-70,000 particles by the end of the
simulation.  The code includes an equation of state valid from
densities below 1\,g\,cm$^{-3}$ up to nuclear densities (including an
estimate of nuclear statistical equilibrium).  Neutrino transport is
followed using flux-limited diffusion neutrino scheme for three
neutrino species (Herant et al. 1994).  The neutrino emission and
cross-sections are also outlined in Herant et al. (1994).

Our simulations are set to mimic the conditions 2-10\,s after the
launch of the explosion when material begins falling onto the
proto-neutron star.  The initial condition begins with a set of
particles ranging from 1000\,km up to 10,000\,km.  The material is
given velocities set to the free-fall velocity (In our 1-dimensional
simulations, the infall velocity is within 1-10
velocity) at their radial position with densities set to give the
desired accretion rate (recall, $\rho=\dot{M}/4 \pi r^2 v_{\rm
infall}$.  With $v_{\rm infall}$ within 1-10\% of the free-fall
velocity, we can accurately set up our initial conditions: our 3
separate rates correspond to 0.001,0.01,0.1\,M$_\odot$\,s$^{-1}$
(Table~\ref{tab:neut}).  For these simulations, we model only constant
fallback rates.  Note that in Nature, the fallback rate varies quite a
bit.  But we are focusing our study on the neutrino emission at peak
fallback rates, and this suite of simulations will definitely bracket
the range of results.  The entropy of the infalling material is
generally set to a few.  The results are fairly insensitive to this
initial entropy as the shock resets the entropy.  Boundary conditions
are probably the biggest uncertainty in our calculations.  Particles
are fed in through the outer boundary and are allowed to accrete
through the inner boundary and be ejected out of the outer boundary.
The infall through the outer boundary is determined by a fixed infall
rate.  This infall is only altered if material is flowing out of this
outer boundary.  At any point where an outflow occurs, the inflow is
temporarily halted.  This models the effect of outflows choking off
the accretion.

The inner boundary is more difficult.  Ideally, we would model the
matter until its density reaches neutron star densities ($\sim 10^{14}
{\rm g cm^{-3}}$) and it is mostly deleptonized.  At such time, the
matter has lost most of its energy and we can be sure that we have
accounted for the total neutrino luminosity.  However, the sound
speeds at such densities and the size of our Lagrangian particles near
the proto-neutron star would decrease the timestep to fractions of a
microsecond.  Such small timesteps prohibit us from following the
evolution of the fallback for more than a fraction of a second.
Instead we opt to use slightly less demanding criteria for the removal
of particles on the inner boundary.  Our standard set of models
accretes particles whose density rises above $\sim 10^{10} {\rm g
cm^{-3}}$ with electron fractions below 0.3.  This means that we are
assuming any further energy released by the matter as it accretes goes
into heating the neutron star which will cool on longer timescales.
In our suite of models, we include a test of this boundary condition
and find that the total neutrino luminosity\footnote{Note, that except
for our low accretion rates, our inner boundary at an neutrino depth
of more than a few (above 10 in the highest accretion rates).  It gets
close to, and at times is below, 2/3 for low accretion rates.} is not
too sensitive to our assumption for accretion.

We assume the axis of rotation lies along our axis of symmetry in our
2-dimensional caclulation.  Each particle is given an angular momentum
and it retains that angular momentum for the duration of the
calculation.  The angular momentum for each particle is determined to
fit the angular velocities of our progenitor stars (Fig. 2), so
material along the axis has a low angular momentum whereas material in
the equator has the highest angular momentum.  The specific angular
momentum $j$ is given by $j=(x/10,000\,km)^2 \omega$ where $\omega$ is
the angular velocity taken from the stellar models.  The inner material
starts below the angular momentum given by $\omega$, but by the end of
our simulations, most of the material near the neutron star surface
has the full angular momentum set by this $\omega$ value.  There is no
angular momentum transport and hence, technically, no heating from
this transport.  However, the angular momentum will slow the materials
inflow, breaking the symmetry in the downflow.  Fryer \& Heger (2000)
also found that the angular momentum alters the instability criterion,
preventing convection in the equatorial region where the angular
momentum gradient is highest.

To study the effect of a hot neutron star, we introduced a non-zero
neutrino flux arising from our inner, neutron-star, boundary.  In
particular, we are interested in how the neutrinos from a hot neutron
star affect the hydrodynamics, so we have chosen rather large
luminosities ($10^{52},2\times10^{52} {\rm ergs \, s^{-1}}$ electron
neutrino luminosities with energies of 10\,MeV with corresponding
$8\times10^{51},1.6\times10^{52} {\rm ergs \, s^{-1}}$ anti-electron 
neutrino luminosities with energies of 15\,MeV).  These neutrinos 
are a boundary source for our flux-limited diffusion transport 
scheme and transported out of the system, heating the inflowing 
material.  However, in most of our simulations, the neutrino 
optical depth is fairly low, and the total energy deposited is 
also minimal (see \S 3.2).

Finally, there is always concern that the artificial viscosity 
used in SPH is introducing spurious effects into our calculations.  
For the most part, our studies have found this numerical artifact 
to play a small role in results studying core-collapse supernovae, 
and we do not expect it to play a large role in these calcualtions.  
Nonetheless, we have included a simulation where we have increased 
both the bulk and von Neumann-Richtmyer viscosities by a factor 
of 2 (3.0,6.0 respectively versus our standard values of 1.5,3.0).

\begin{figure}
\plotone{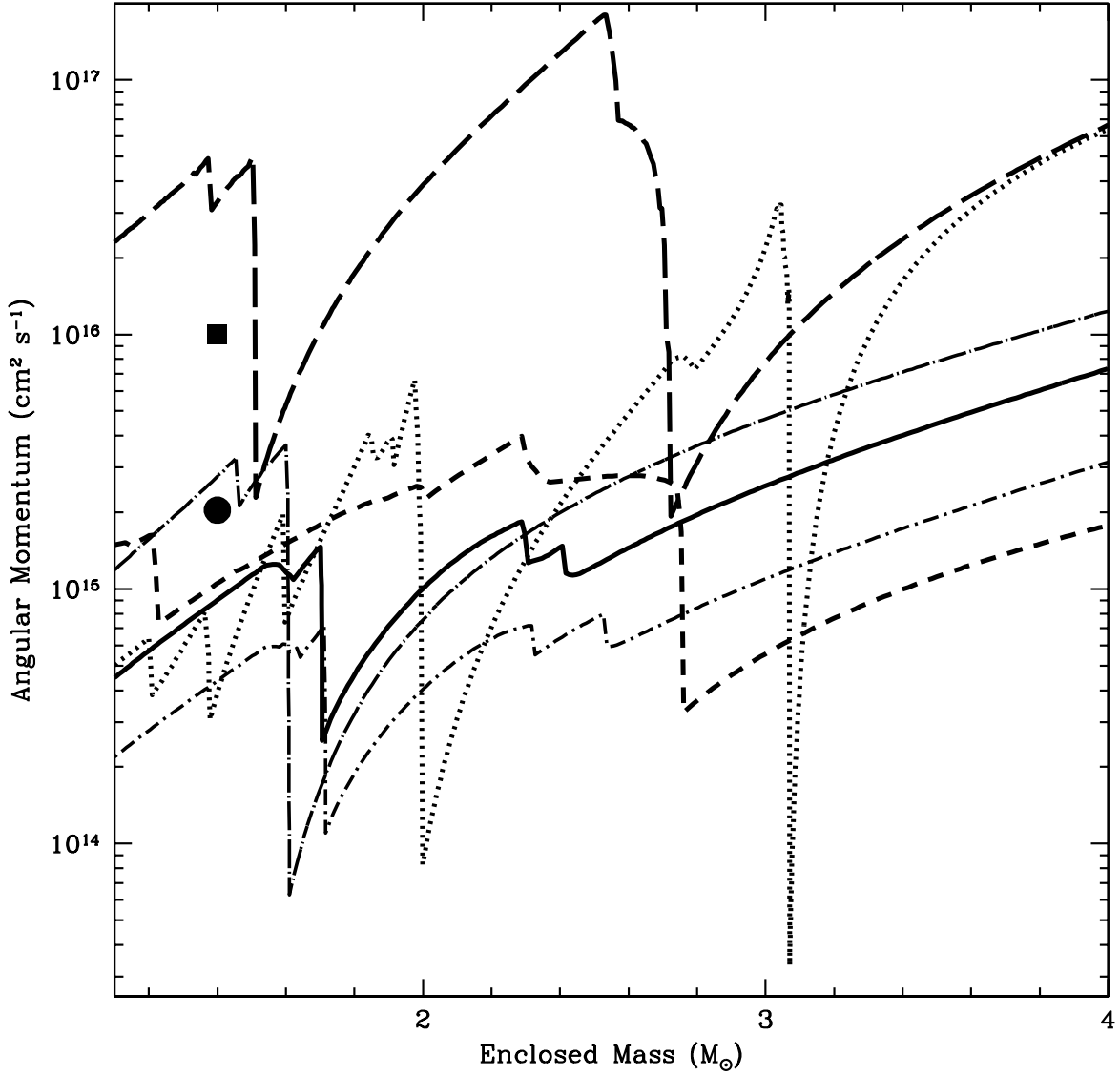}
\caption{Angular momentum versus enclosed mass for 6 different stellar
models: standard 20\,M$_\odot$ model (solid), standard 12\,M$_\odot$
model (dotted), standard 35\,M$_\odot$ model (dashed), 20\,M$_\odot$,
$N^2_{\mu}=0.1$ model (short-dot dashed), 20\,M$_\odot$
$B_{\mu},B_{\phi}=0.1$ model (long-dot dashed), and a
100\,km\,s$^{-1}$ model with no magnetic braking.  The first 5 models
are from Heger et al. (2005), the last model is from Heger et
al. (2000).  The angular momenta used in our simulations is shown by
the circle (standard) and square dots.}
\label{fig:1drot}
\end{figure}

With both numerical and physical effects to study, we have run a suite
of simulations to test the dependence of the neutrino luminosity on
the initial conditions and on the numerics.  A summary of this suite
of models, along with their basic results, is summarized in
table~\ref{tab:neut}.  Our base model has an accretion rate of
0.01\,M$_\odot$ s$^{-1}$ with an angular momentum equal to that shown
in the circle in Figure~\ref{fig:1drot}.  We also include higher
rotating runs with angular momenta set by equating the angular momenta
to the value denoted by the square in Figure~\ref{fig:1drot}.  In both
cases, the angular momentum is low, so we don't expect the formation
of a full accretion disk.  We study the inner boundary in a number of
ways.  We include a simulation where the criteria for accretion is
more strict: $\sim 10^{11} {\rm g cm^{-3}}$ with electron fractions
below 0.1.  We also include a set of simulations with an absorbing
boundary at 100\,km (a black hole boundary condition).  We study
accretion rates 10 times higher and ten times lower than our canonical
rate.  Finally, since the accretion occurs at early times, we have
also included a few simulations where the neutron star itself is still
emitting neutrinos.  In \S~\ref{sec:dynamics}, we compare the results
on the dynamics for this suite of calculations.  In
\S~\ref{sec:neutrinos} we compare the resulting neutrino
luminosities.

\subsection{Accretion Dynamics\label{sec:dynamics}}

\begin{figure}
\plotone{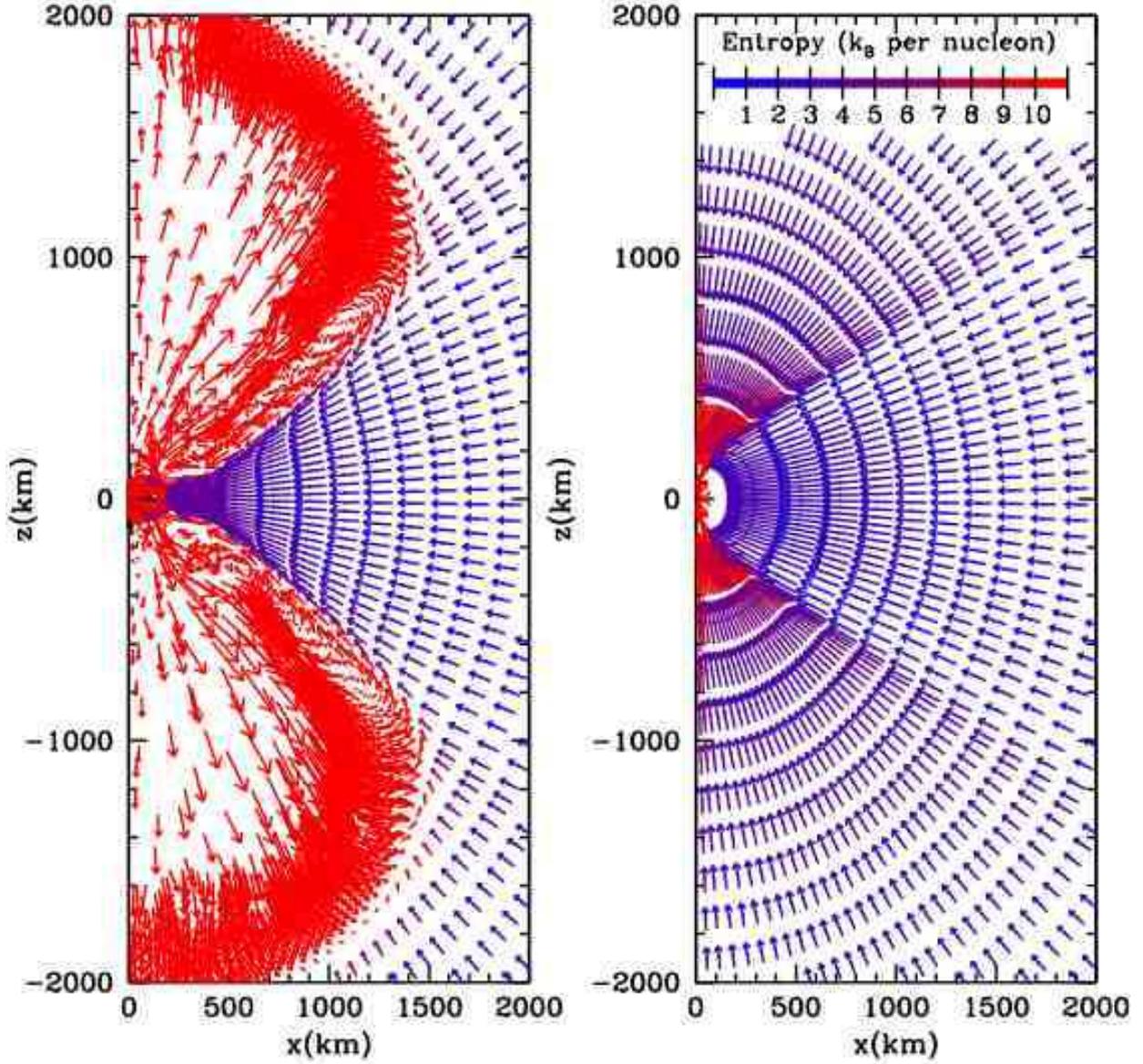}
\caption{Particle position from our 2-dimensional smooth particle
hydrodynamics calculations shaded by entropy.  The direction and
length of the vectors denote velocity magnitude and direction.  The
left panel shwos the results at 0.15\,s from our standard neutron star
``NS2'' model.  The shocked, high-entropy, material rises and drives
an outflow.  The right panel shows the absorbing boundary ``BH2'' 
model at the same time.  Note that the angular momentum alters the 
flow, but does not slow the material in the equator enough to 
produce enough viscous heating to drive outflows.}
\label{fig:comp-nsvbh}
\end{figure}

Before we discuss the neutrino fluxes from this suite of calculations,
let's compare the dynamics in the accretion.  We expect material to
shock as it hits the proto-neutron star, and some of this shocked 
material will begin to rise, driving convection.  These simulations 
are 2-dimensional.  Even without angular momentum, perturbations 
along our symmetry axis would allow the instability to develop 
stronger along this axis.  But we have studied this in some 
detail in core-collapse calculations and have minimized this 
effect (see, for example, Fryer \& Heger 2000).  However, the 
fact that the angular momentum axis also lies along the symmetry 
axis drives convection along our axis of symmetry.  The growth 
of convective instabilities is stabilized by angular momentum 
gradients and the deceleration of material (along with the fact 
that our initial condition has an angular momentum gradient 
which is strongest along the equator), the convection initially 
grows strongest in the poles.  Figure~\ref{fig:comp-nsvbh} 
shows the results of our standard calculation showing this outflow.  This 
figure shows the evolution at 0.15\,s.  The right panel shows the evolution 
of the corresponding ``black hole'' simulation with the absorptive boundary.  
The true innermost stable circular orbit for a slowly rotatign 
3\,M$_\odot$ black hole is closer to 20\,km, so our 100\,km absorption 
radius understimates the activity from a real black hole.  However, 
we can see that although the angular momentum alters the inflow, it 
is insufficient to stop it, and the material continues to accrete 
directly onto the black hole.

\begin{figure}
\plotone{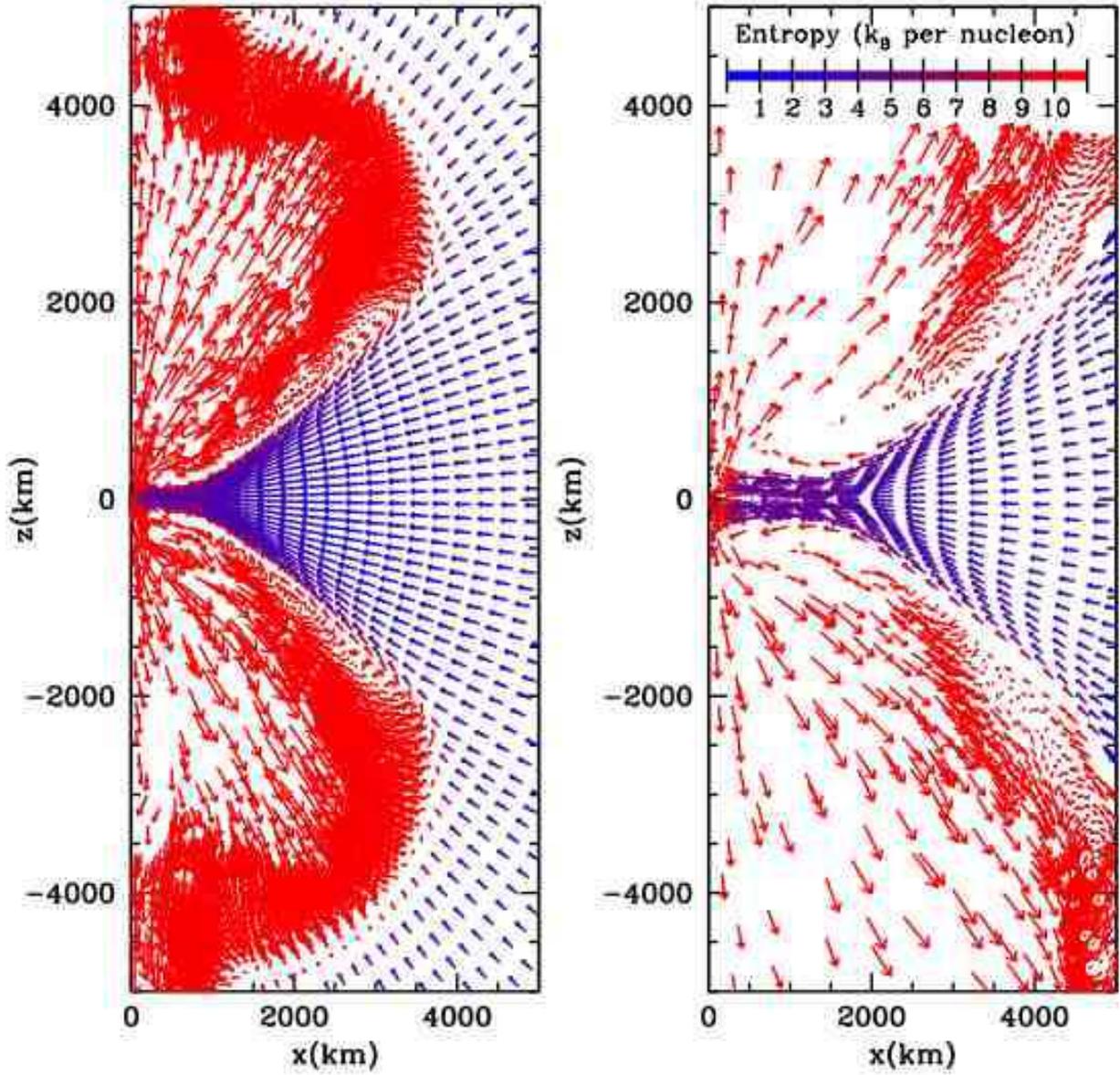}
\caption{Same as Fig.~\ref{fig:comp-nsvbh} but for the standard 
NS2 model 0.3 and 0.45\,s from the start of the simulation.  Note 
that the axis has been extended to show the extent of the outflow.}
\label{fig:comp-late}
\end{figure}

As our neutron star model evolves the outflow expands, constraining
the accretion to a funnel roughly $45^{\circ}$ from the plane.
Figure~\ref{fig:comp-late} shows the evolution of our standard model
to 0.3 and 0.45\,s.  Over half of the inflowing material is ultimately
ejected.  Not only does the ejected material not contribute to the
energy available for neutrino emission, some of the energy of that
accreted material goes toward accelerating this ejecta.  

Such outflows have been studied for over a decade in supermassive
black hole systems such as active galactic nuclei (see Blandford \&
Begelman 1999) for a review.  The amount of potential energy released
in the accretion of material onto a compact object is enormous.  If
this energy is tapped (either by viscous forces or neutrino emission),
it can drive an explosion.  Rockefeller et al. (2007) and Fryer et
al. (2006) have recently applied this physics to stellar-massed
compact objects to better understand their simulations of accreting
systems.  In rotating black hole systems, material hangs up in a disk.
Viscous forces that transport out angular momentum also transport
energy, driving outflows.  What prevents this ejection is cooling
(either via photon radiation in the case of supermassive black holes
or neutrinos in the case of most collapsing systems).  Of course, for
black hole systems that do not have a hard surface preventing the
inflow of material, high angular momentum is required to prevent the
energy from all flowing directly into the black hole.  In neutron star
systems, outflows (or at least vigorous convection) have been expected
for more than a decade (Chevalier 1989; Fryer et al. 1996).

\begin{figure}
\plotone{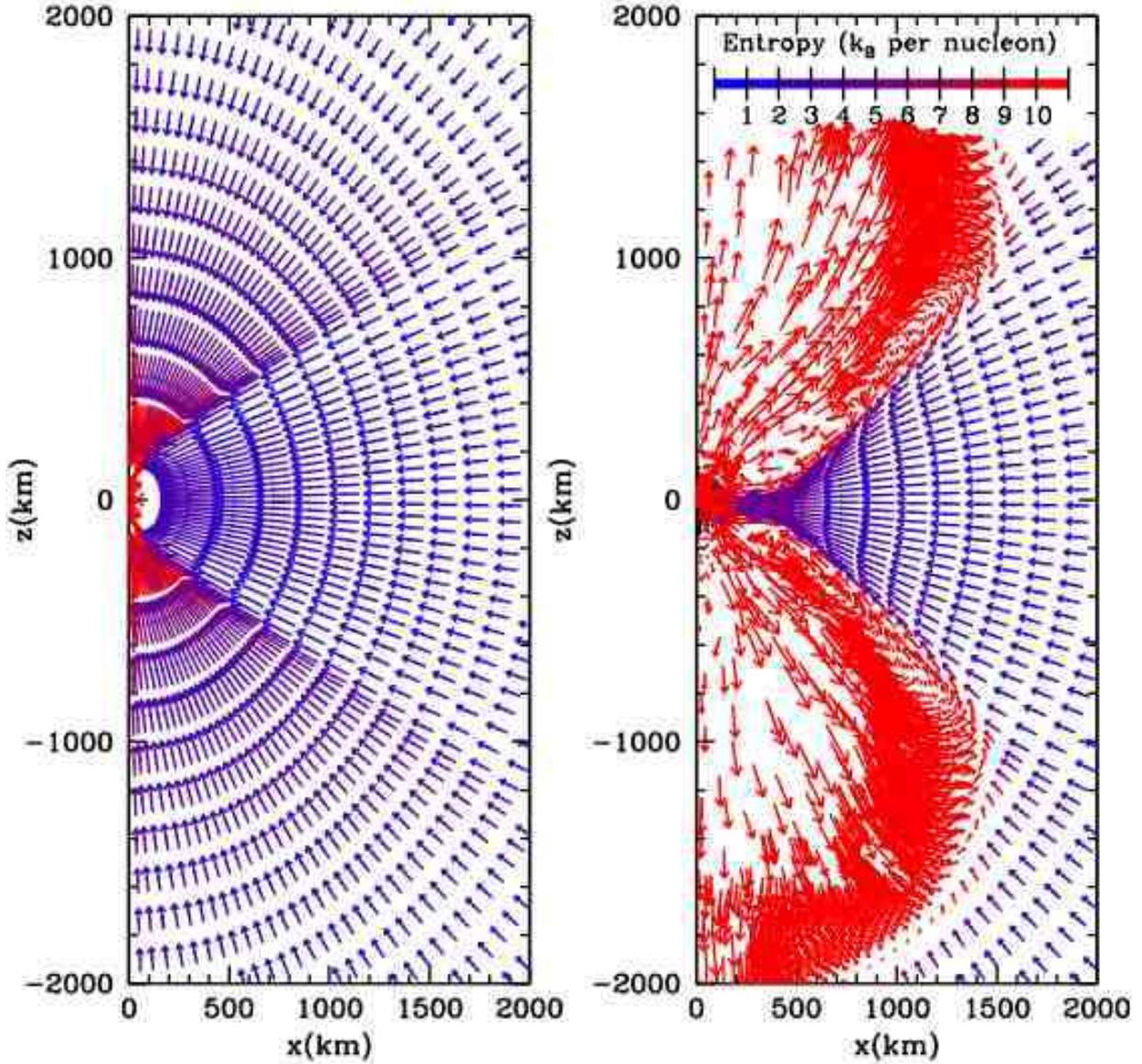}
\caption{Same as Fig.~\ref{fig:comp-nsvbh} but for the two
faster-rotating models: BH2-Rot10 (left panel) and NS2-Rot2 (right
panel).  A factor of 10 increase in the square of the specific angular
momentum ($j^2$) is insufficient to form a disk in our models and the
dynamics is not changed significantly.}
\label{fig:comp-rot}
\end{figure}

If the angular momentum were high enough, the infalling material would
hang up in a disk.  This is most important in the black hole systems.
We ran a series of models increasing the specific angular momentum
($j$) a factor of 3 and, for our black hole models, a factor of
10.  MacFadyen \& Woosley (1999) found that the specific angular
momentum must be at least $\sim 10^{17} {\rm cm^2 s^{-1}}$ (a factor
of 1000 higher than our standard $j^2$ value).  As we can see from
Figure~\ref{fig:comp-rot}, our factor of 10 increase in $j$ is not
enough to change the fate of material falling back on our ``black
hole'' simulation with its large absorptive boundary.  In a true black
hole system, the angular momentum available would increase as we move
beyond 3\,M$_\odot$ (well above the angular momenta shown in
Fig.~\ref{fig:1drot}).  This is one reason why the collapsar GRB model
argues for systems where the black hole mass exceeds 3\,M$_\odot$ when
the angular momentum in the star is sufficient to produce an accretion
disk.  Neutrinos from these systems have been considered in detail
elsewhere (MacFadyen \& Woosley 1999, Hungerford et al. 2006,
Rockefeller et al. 2007) and we do not study them in this paper.

\begin{figure}
\plotone{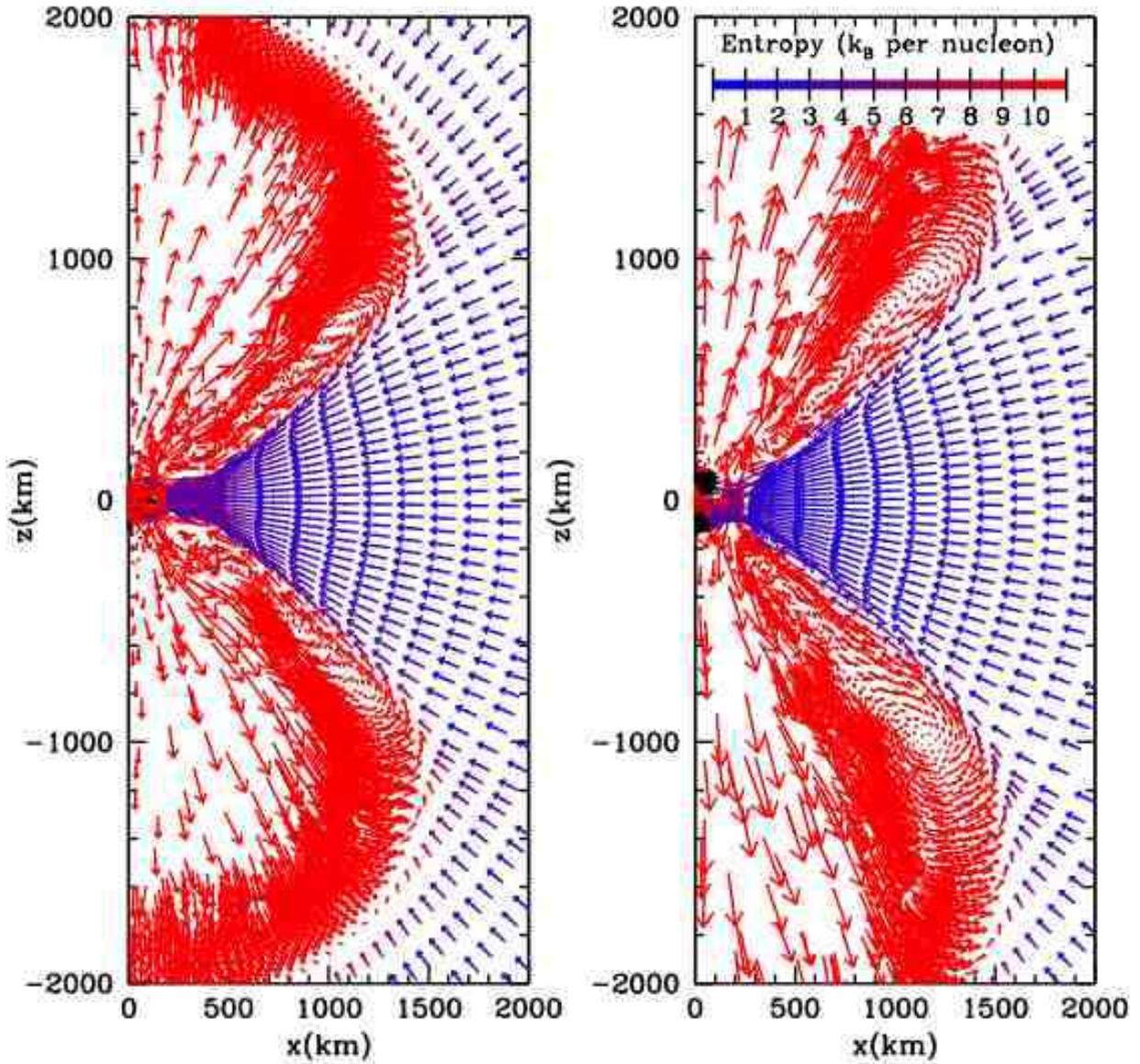}
\caption{Same as Fig.~\ref{fig:comp-nsvbh} but for two of our 
models using hot neutron stars:  NS2-Hot1, NS2-Hot2.  The 
dynamics of these simulations is very similar to our cold neutron 
star runs, but the explosions are slightly stronger.}
\label{fig:comp-hot}
\end{figure}

If the neutron star is still hot and emitting neutrinos, it can also
alter the inflow of fallback.  Figure~\ref{fig:comp-hot} shows two
simulations with varying amounts of neutrino energy (18 and
36$\times10^{51} {\rm ergs \, s^{-1}}$) arising off the neutron star
surface.  The effect on the dynamics is minimal, although a slight
increase in the velocity (and hence position at a given time) can be
seen.  As we shall see below, the neutrinos emanating from the
proto-neutron star surface will dominate the total neutrino flux in
such cases.  But we don't expect these high luminosities at 10\,s 
and the accretion luminosity will dominate at these later times.  
But at early times, the modification of the downflow on the 
neutrino opacities is the dominant effect.  

\begin{figure}
\plotone{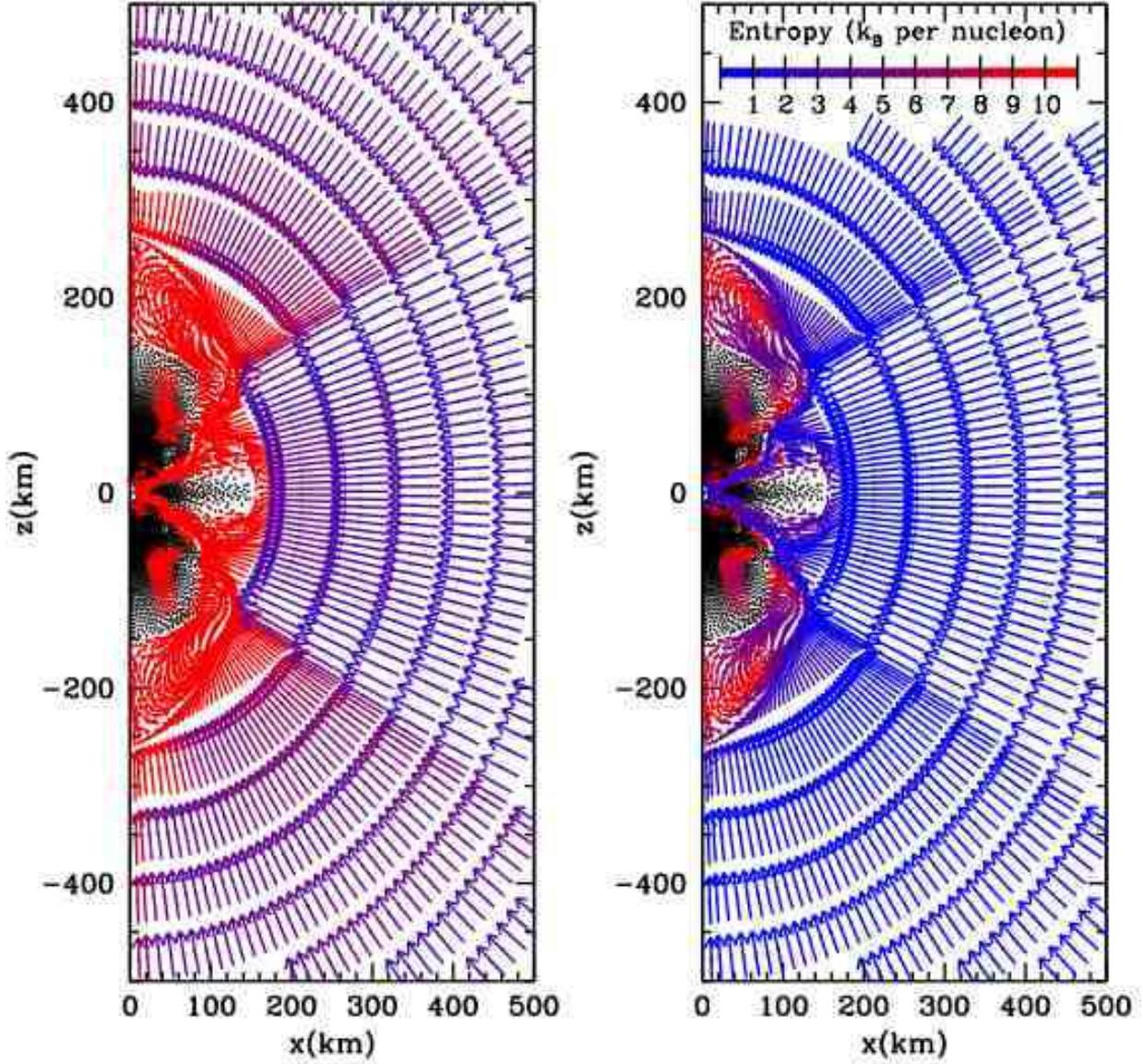}
\caption{Same as Fig.~\ref{fig:comp-nsvbh} but for a neutron star 
models with accretion rates of 0.001 and 0.1\,M$_\odot$\,s$^{-1}$.}
\label{fig:comp-mdot}
\end{figure}

Finally, we have varied the infall rate from 0.001 to
0.1\,M$_\odot$\,s$^{-1}$.  The effect this has on the dynamics is
shown in Figure~\ref{fig:comp-mdot}.  As with many of our models, 
the nature of the dynamics is not altered significantly by these 
changes.  But we shall see that the neutrino flux very much depends 
on the accretion rate.

\subsection{Neutrino Emission\label{sec:neutrinos}}

\begin{figure}
\plottwo{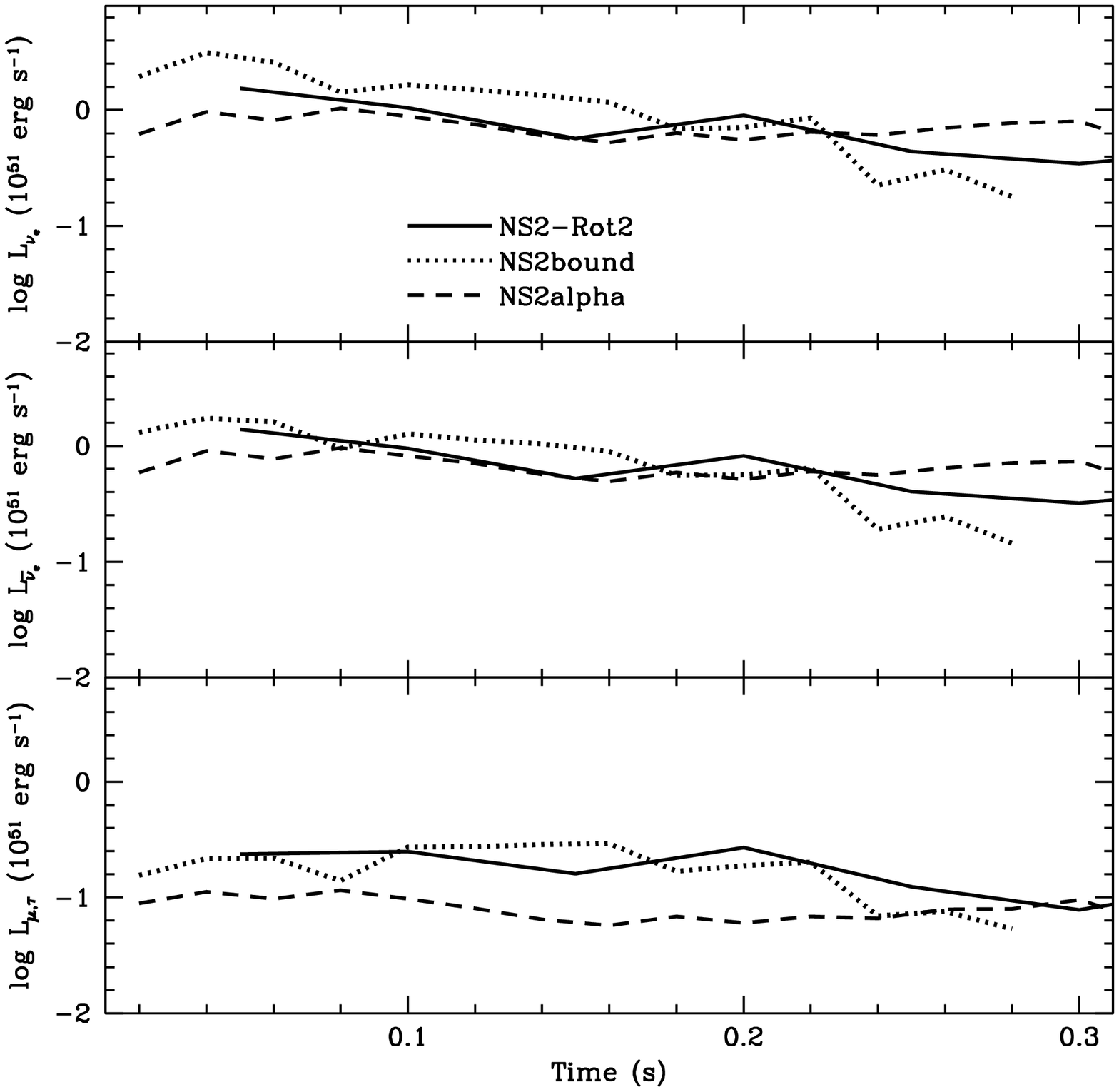}{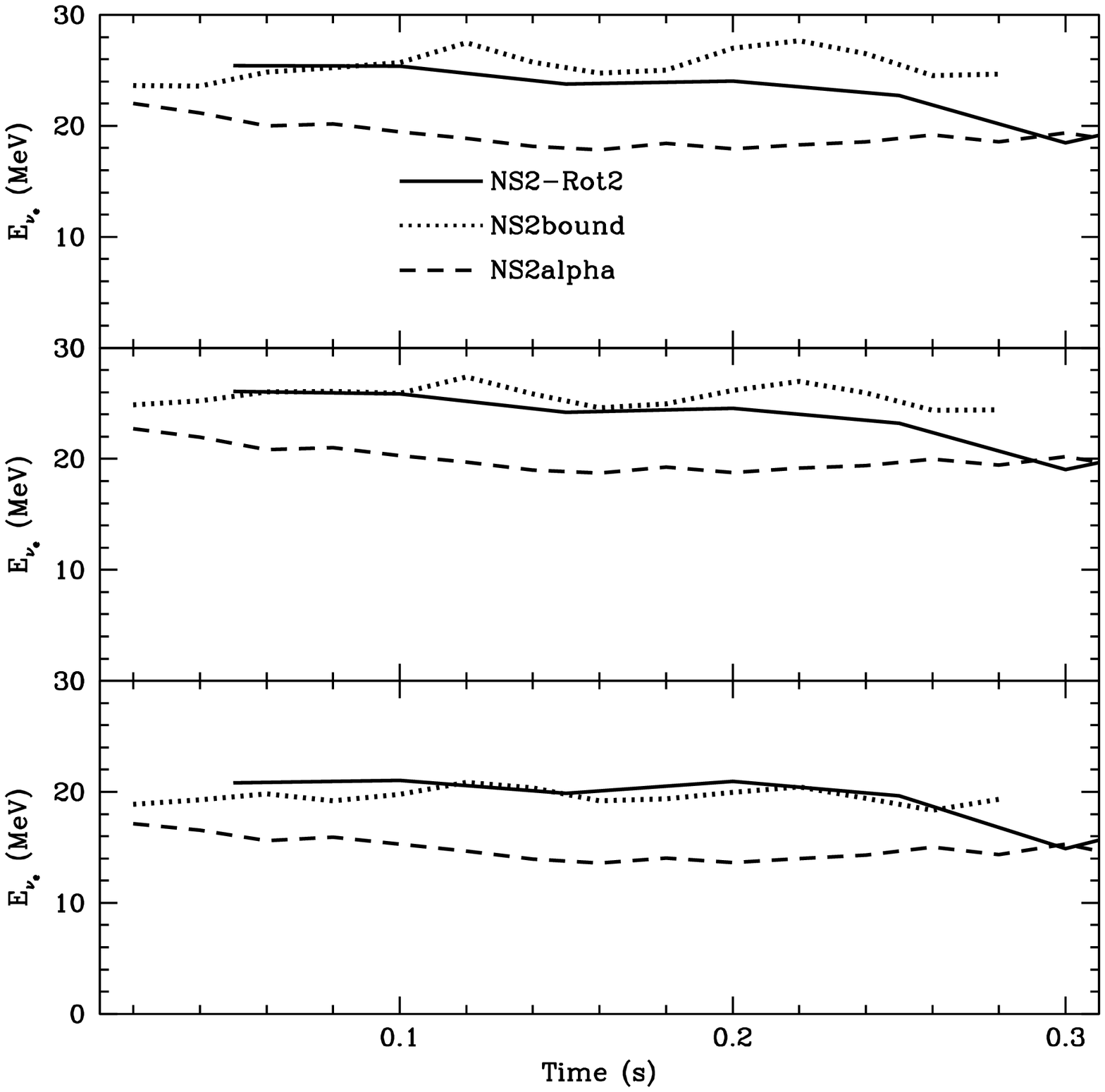}
\caption{Neutrino Luminosity (left panel) and energy (right panel)
versus time for electron, anti-electron and $\mu,\tau$ neutrinos as a
function of time for 3 different models: NS2-Rot2, our fast rotating
NS model (solid line), NS2bound, the simulation with more restrictive
accretion criteria (dotted line), and NS2alpha, the simulation with
the enhanced values for viscosity (dashed line).  Not that all models 
agree to within a factor of 2 in the lumosity and 5\% in neutrino 
energy at all times.}
\label{fig:nunum}
\end{figure}

We found in the last section that the dynamical behavior of the
fallback was relatively insensitive to both uncertainties in the
numerics as well as uncertainties initial conditions: e.g.  fallback
rate and rotation.  But what about the neutrino luminosity?  First
let's study the uncertainties in the numerics.  Fig~\ref{fig:nunum}
shows the neutrino luminosity and mean energy for 3 different models
studying the numerical effects (both the artificial viscosity and the
treatment of the inner boundary) on the calculation.  The variations
in the viscosity and the inner boundary lead to differences that are
less than a factor of 2 in the neutrino luminosity and 5\% variations
in the neutrino mean energy.  Many of these errors could be dominated
by the explicit transport scheme used in these calculations and it is
possible that these errors can be significantly diminished with 
implicit schemes that are more stable.

\begin{figure}
\plottwo{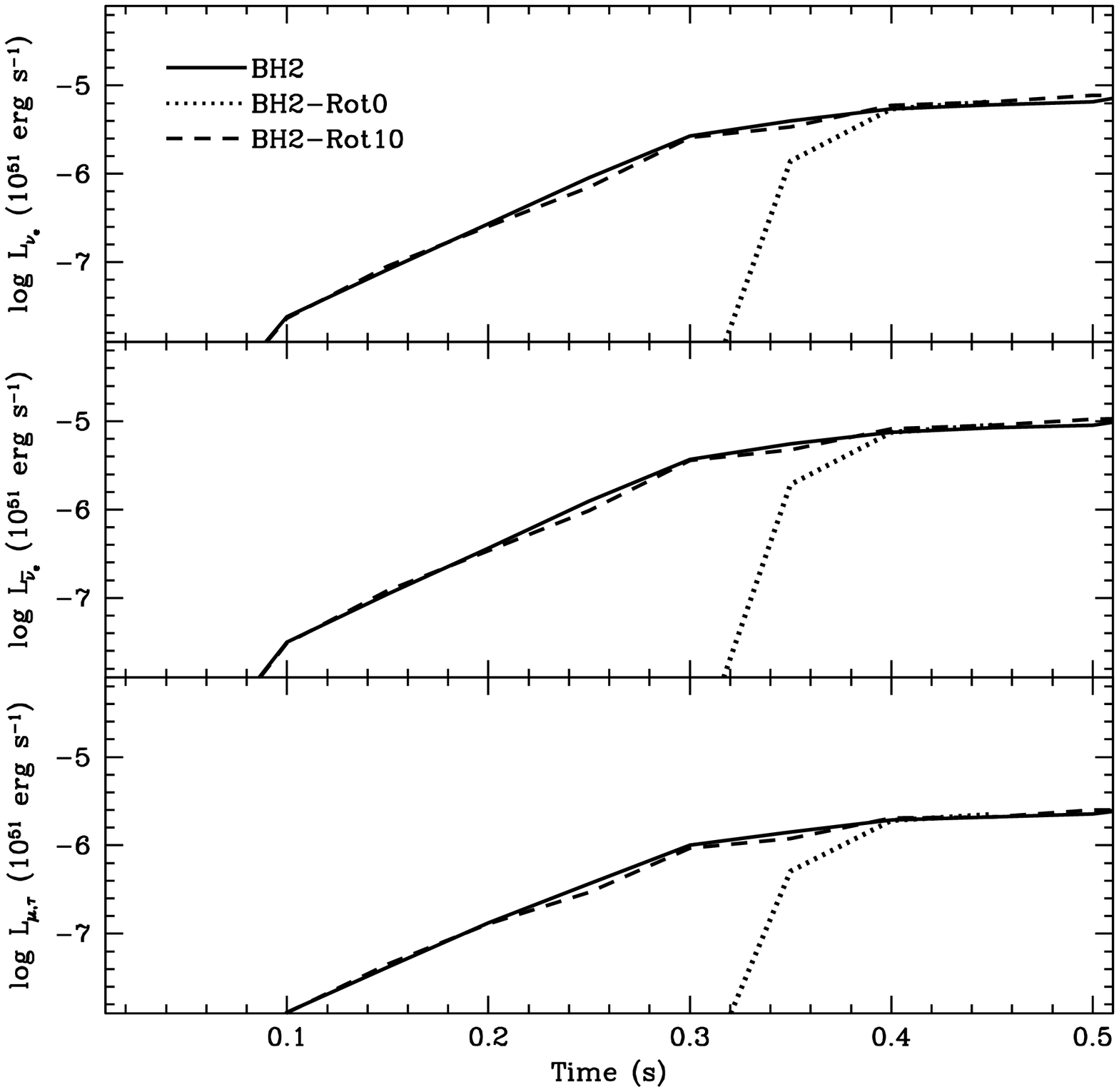}{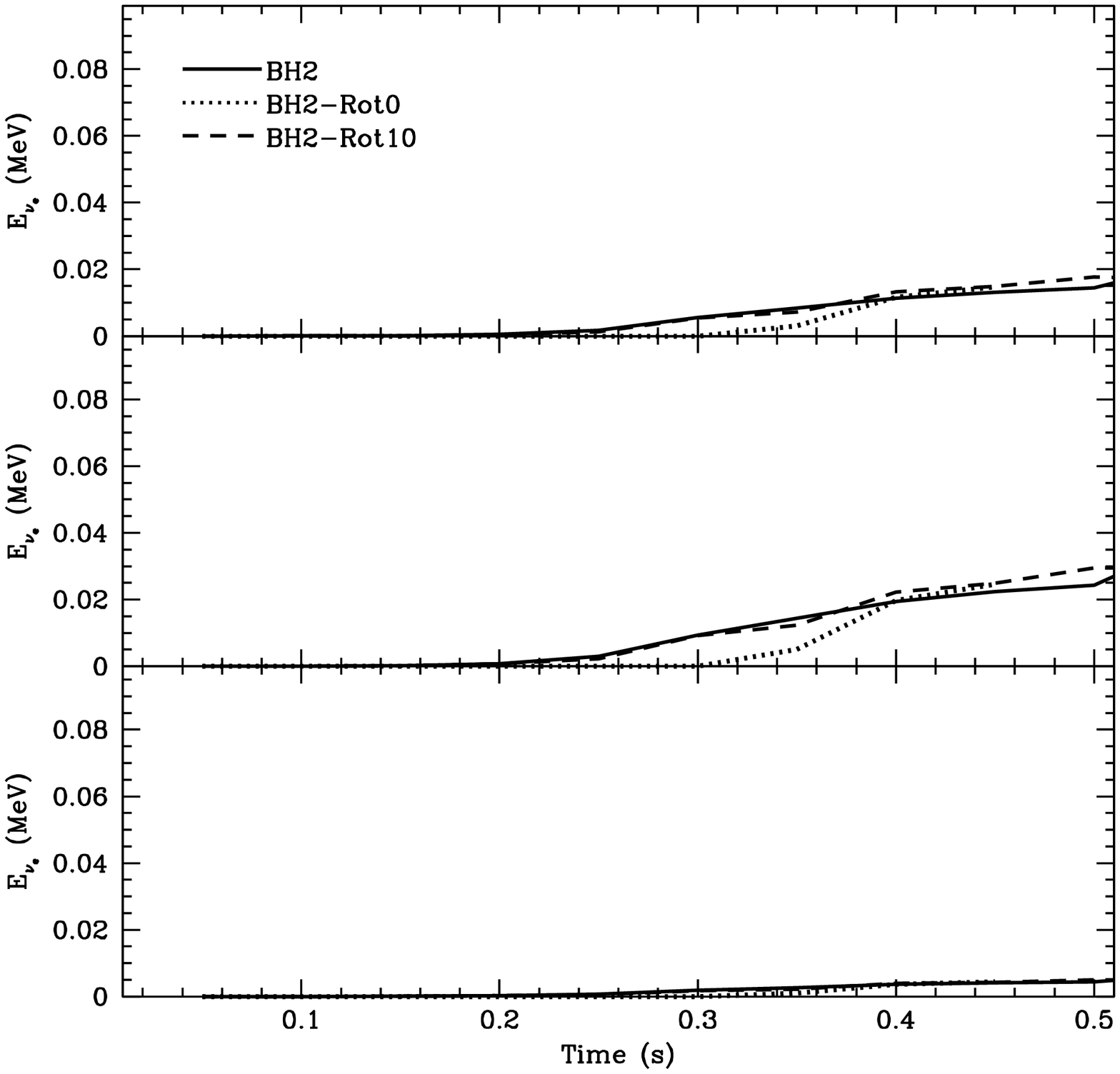}
\caption{Neutrino Luminosity (left panel) and energy (right panel)
versus time for electron, anti-electron and $\mu,\tau$ neutrinos as a
function of time for 3 different black hole models:  BH2 (solid), BH2-Rot0 
(dotted),. BH2-Rot10 (dashed).  These results confirm that the accreting 
material with such low angular momenta falls directly onto the black 
hole without any appreciable neutrino emission.}
\label{fig:nubh}
\end{figure}

Figure~\ref{fig:nunum} also shows the results for a fast rotating NS
model (NS-Rot2).  With the low angular momenta in our calculations
based on the rotation velocities in the inner core material of massive
stars (Fig.~\ref{fig:1drot}), rotation does not alter the neutrino
luminosity noticeably.  Black hole systems depend more sensitively on
the rotation because it is the angular momentum that prevents the
material from accreting directly into our black hole.  But as we
expected from the fact that our angular momentum is too low to produce
accretion disks, the neutrino emission from our black hole systems is
negligible (Fig.~\ref{fig:nubh}).  In such low-angular momentum
systems, we expect essentially no emission after the collapse of the
neutron star down to a black hole.  Contrast this to typical collapsar
conditions, where the material falling back onto the black hole has
enough angular momentum to hang up in a disk and its falback accretion
rate is high enough to produce high-density, high-temperature
structures.  In this case, the neutrino emission from a black hole can
be quite large - with luminosities on par with the neutrino burst of
the original collapse (MacFadyen \& Woosley 1999; Hungerford et
al. 2006; Rockefeller et al. 2007).

In our models, the $\mu$ and $\tau$ neutrinos tend to be an order of
magnitude lower than the corresponding electron neutrino fluxes.  So
our assumption that most of the neutrinos are emitted as electron or
anti-electron neutrinos is reasonably valid.  Hungerford et al. (2006)
and Rockefeller et al. (2007) found that for collapsar models, the
$\mu$ and $\tau$ neutrinos make up a sizable fraction of the total
emission.  At higher accretion rates (and higher angular momentum in
the infalling material), the fraction of the luminosity emitted in 
$\mu$ and $\tau$ neutrinos will likely increase.

\begin{figure}
\plottwo{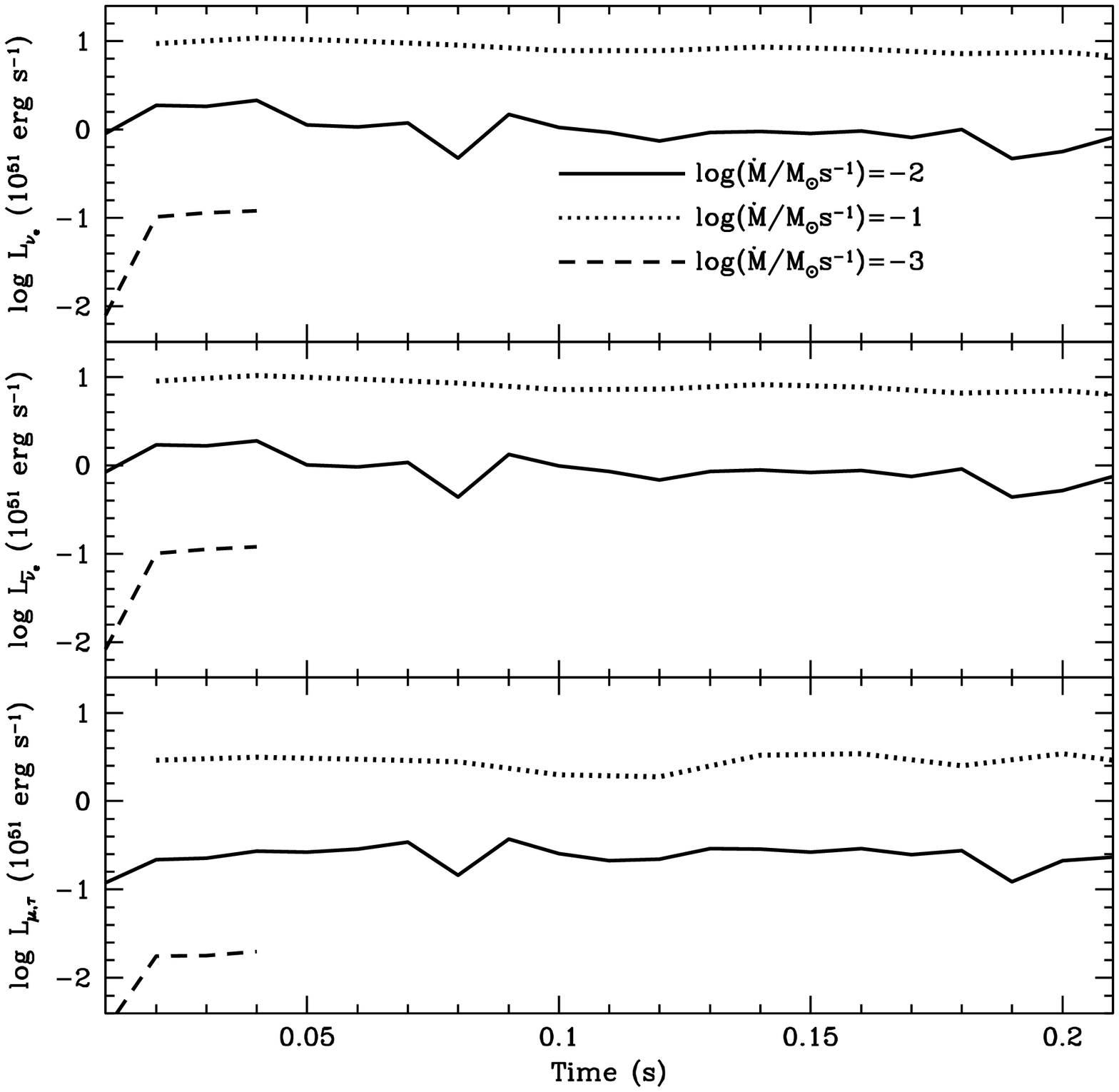}{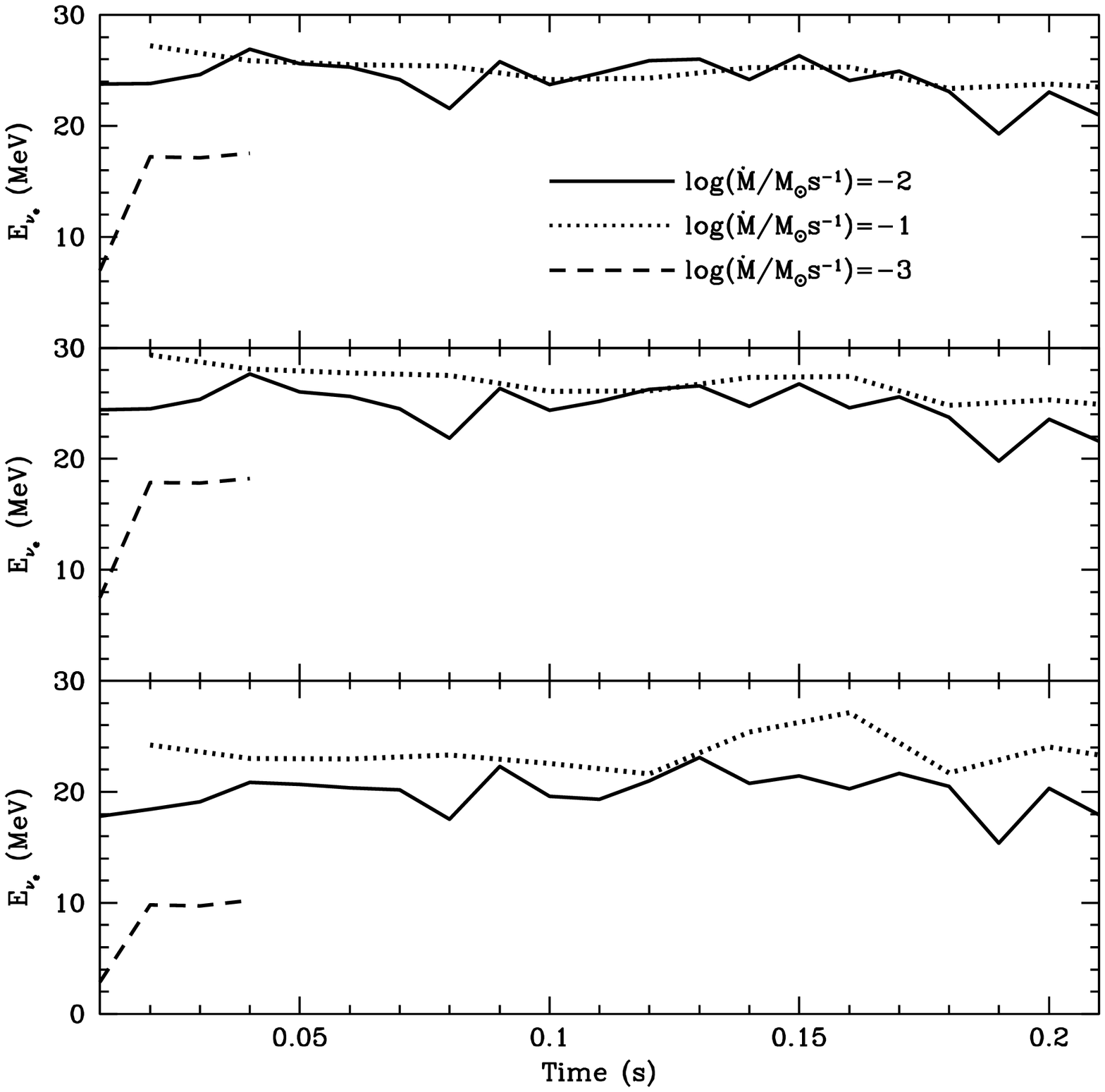}
\caption{Neutrino Luminosity (left panel) and energy (right panel)
versus time for electron, anti-electron and $\mu,\tau$ neutrinos as a
function of time for 3 different accretion rates: 0.001-NS3 (dashed), NS2 0.01
(solid), and NS1 0.1 (dashed) \,M$_\odot$\,s$^{-1}$.  Here the results depend
nearly linearly on the accretion rate and for the highest accretion
rates, the neutrino luminosity is roughly $10^{52}$\,ergs\,s$^{-1}$.}
\label{fig:numdot}
\end{figure}

The strongest dependency on our initial conditions in the mass
accretion rate.  Figure~\ref{fig:numdot} shows the neutrino emission
from 3 different accretion rates onto our neutron star surface.  Our
basic potential energy released estimate would argue that the neutrino
luminosity scales with the accretion rate.  And this trend is
basically true for our simulations.  However, note that the electron
and anti-electron neutrinos for our highest accretion rate
(0.1\,M$_\odot$\,s$^{-1}$) case are not quite an order of magnitude
higher than our standard accretion rate (0.01\,M$_\odot$\,s$^{-1}$).  
This is because the electron neutrinos become trapped in this highest 
accretion rate case, lowering the escaping luminosity.  The $\mu$ and 
$\tau$ remain nearly an order of magnitude higher as they are not 
trapped in any of our simulations.

\begin{figure}
\plottwo{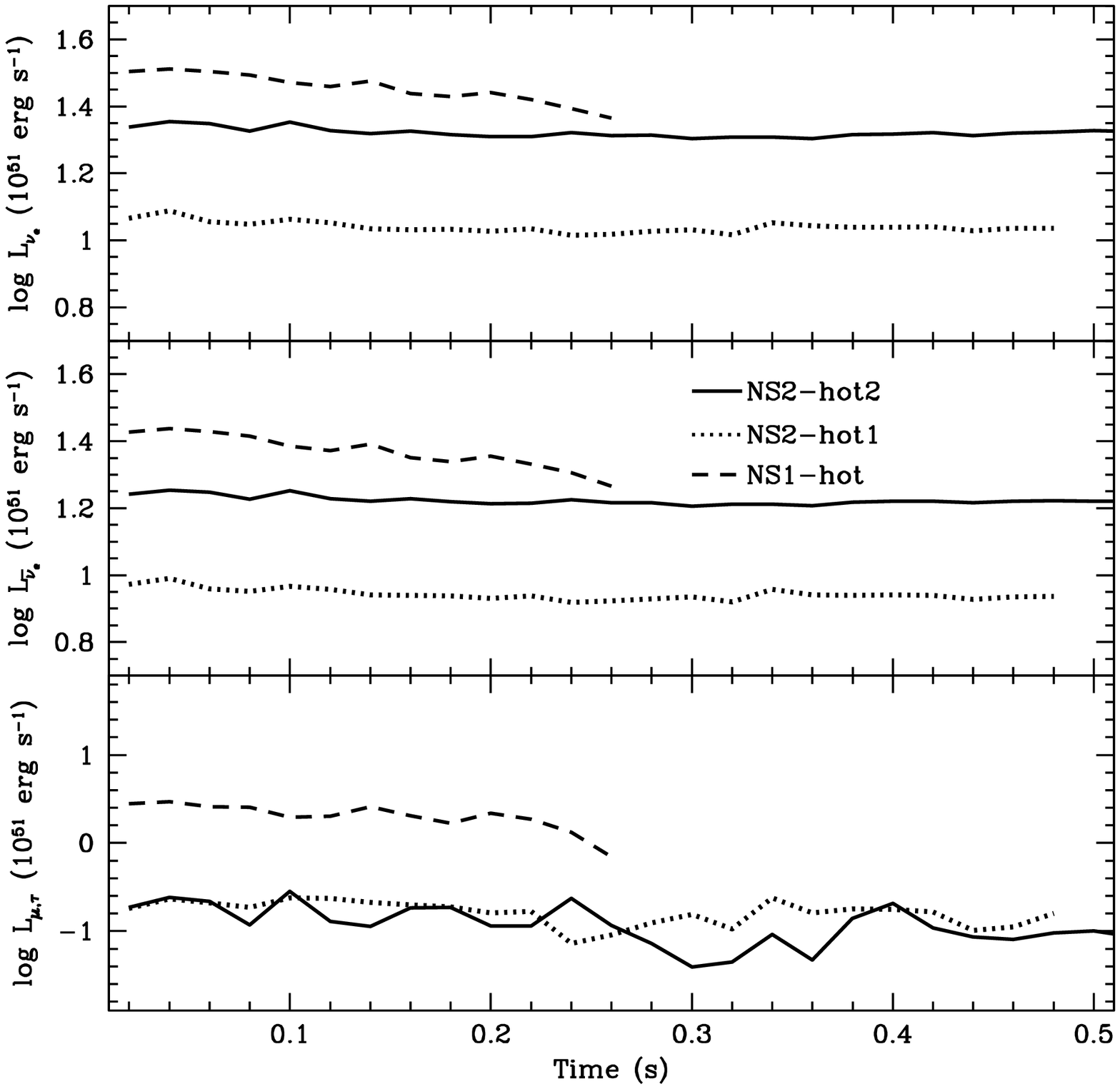}{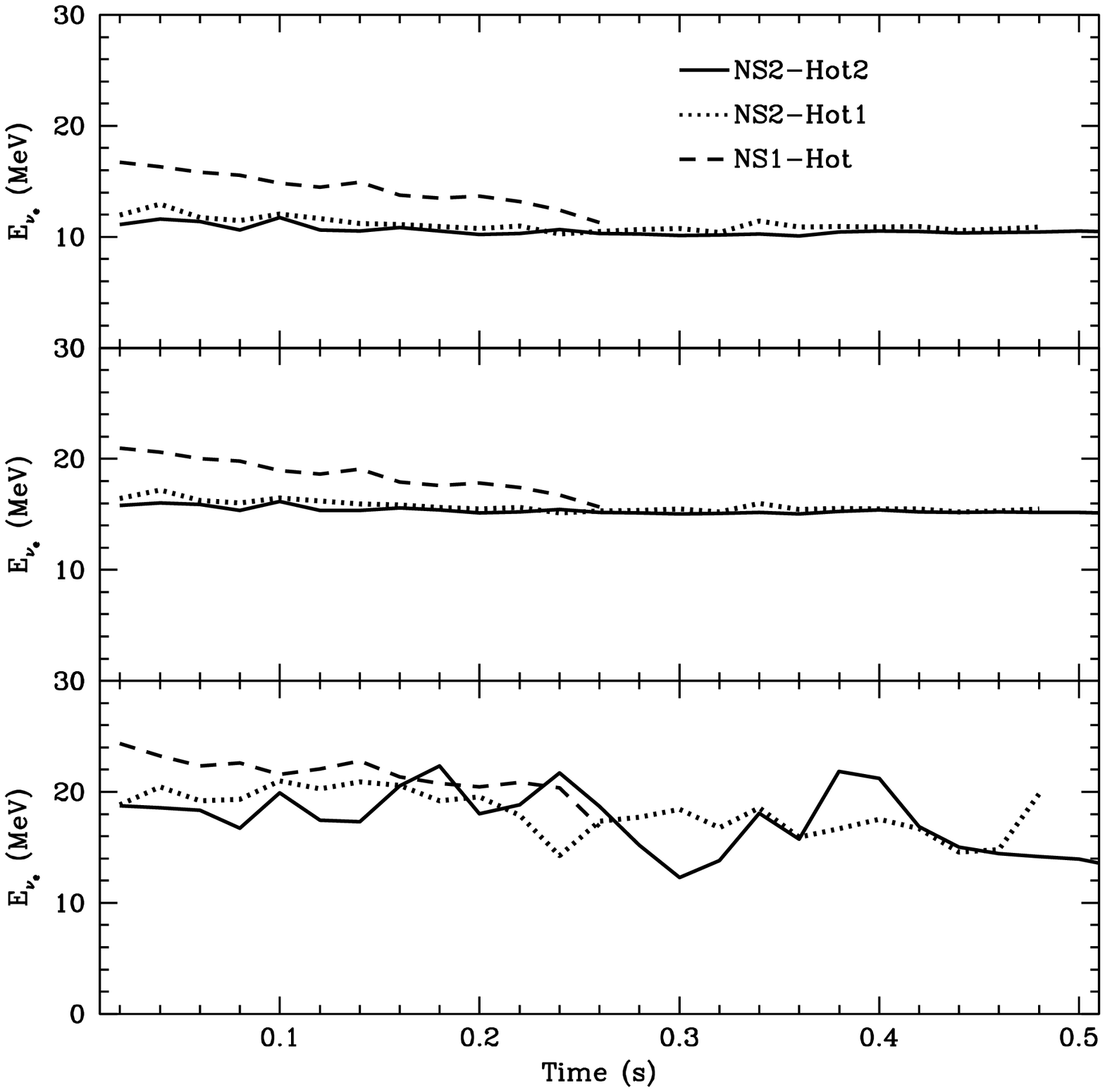}
\caption{Neutrino Luminosity (left panel) and energy (right panel)
versus time for electron, anti-electron and $\mu,\tau$ neutrinos as a
function of time for 3 different neutron star models with emitting
neutron stars: NS2-hot1 (dotted), NS2-hot2 (solid), NS1-hot (dashed).
Fallback contributes an additional 10-20\% of the luminosity on
average for the NS2-hot2 model, 20-40\% to the NS2-hot1 model, and
over 50\% to the NS1-hot model.  The neutrino energy is also altered 
by an amount comparable to the change in the luminosity.  Although a 
hot neutron star may dominate the neutrino luminosity, fallback 
clearly can contribute a sizable fraction of the observed neutrino 
flux.}
\label{fig:nuhot}
\end{figure}

Figure~\ref{fig:nuhot} shows the neutrino emission for 3 different
``hot'' neutron star models.  Fallback contributes an additional
10-20\% of the luminosity on average for the NS2-hot2 model, 20-40\%
to the NS2-hot1 model, and over 50\% to the NS1-hot model.  The
neutrino energy is also altered by an amount comparable to the change
in the luminosity.  Although a hot neutron star may dominate the
neutrino luminosity, fallback clearly can contribute a sizable
fraction of the observed neutrino flux.

Finally, note that in general, our predicted fallback neutrino 
energies are high.  The most notable exception is that the mean 
neutrino energy is 5-10\,MeV cooler for our low accretion-rate
simulations.  As the accretion rate decreases, so too will the 
mean energy of the emitted neutrinos.

\section{Conclusions}

Fallback (of at least 0.1\,M$_\odot$) is likely to occur in normal
($10^{51}$\,erg) supernova explosions from stars with initial masses
of 12\,M$_\odot$ or greater.  With energy-injected (more realistic
than piston-driven) explosion calculations, this fallback occurs
quickly, generally in the first 15\,s and peak accretion rates in the
0.01-0.1\,M$_\odot$\,s$^{-1}$ range should be expected.  If the 
engine stays active long after the launch of the supernova shock, 
the total fallback will be lower.  But neutrino engines weaken 
significantly quickly after the shock is launched and the material 
above the neutrinosphere (absorbing the energy) is ejected.  
Magnetic-driven explosions could be very different.  The mechanism for
this fallback is essentially that proposed by Colgate (1971) whereby
the outflowing material decelerates as it pushes against the material
above it, ultimately decreasing its velocity below the escape velocity
and causing it to fall back onto the proto-neutron star.  This means
that the bulk of the fallback is fairly insensitive to the structure
on the outer envelope of the star (so fallback is roughly the same 
whether or not the star is in a binary).

Our multi-dimensional simulations of this fallback suggest that not 
all of this fallback material is actually incorporated into the 
proto-neutron star.  Some of it flows out of the system.  The 
corresponding neutrino luminosity is also lower, both because less 
material is accreted and some of the energy released goes toward driving 
outflows.  But the neutrinos from fallback could still dominate the 
neutrino emission from a proto-neutron star 10\,s into the explosion.

At this time, observations of neutrinos from supernovae are limited to
those of supernova 1987A (Hirata et al. 1987, Bionta et al. 1987).
With only 20 neutrinos over $\sim$15\,s, it is difficult to place many
constraints on the fallback.  Some authors have claimed that claimed
that the late-time neutrinos could not be easily explained by a
cooling neutron star (e.g. Suzuki \& Sato 1987), but others have
argued that the observed neutrino signal is consistent with neutrinos
diffusing out of a hot proto-neutron star (Burrows \& Lattimer 1987;
Bruenn 1987) .  In our neutron-star forming models, fallback ends
within the first 10-15\,s, comparable to the duration of the observed
neutrino burst and the flux is consistent with our more standard
accretion rates.  The observations are consistent with fallback, but
could easily be explained by a neutron star without any fallback.
Given that the progenitor star is believed to be greater than
15\,M$_\odot$, fallback is likely to have occured.  But with the
current errors in the time-dependent luminosity, it is difficult to
determine whether or not accretion is taking place.

If we assume fallback accretion did occur, the fact that the mean
neutrino energy appears to be dropping with time in the observations
suggests that the accretion rate is dropping at 10\,s.  But neutrinos 
are still observed at 10\,s.  Unless the fallback material has 
considerable angular momentum, the compact remnant is not a black 
hole at 10\,s.  In addition, because the accretion rate is dropping 
dramatically at this time, it is unlikely that the remnant will 
accrete much additional mass.  It will remain a neutron star.  In 
our fallback calculations, systems that accrete enough material 
to form a black hole are either already a black hole at 10\,s or 
still accreting rapidly at 10\,s, neither of which is supported by 
the observations.  But we really need a better neutrino signal to 
say anything definitive.

With such high accretion rates, fallback can easily dominate the
neutrino luminosity after a few seconds, more than doubling the
emission from the neutron star at thearly times.  If we wish to study
neutrino opacities in the supernova explosion, we will have to be able
to calculate this fallback.  It may be possible to estimate the
fallback rate from detailed study of supernova light-curves.
Fortunately, any system that is detected in neutrinos will have a
wealth of data in photons of all wavelengths.

If the fallback rate is high, the infalling material will alter the
position of the neutrinosphere, and the problem becomes nearly as
complex as the supernova explosion modeling itself.  We have only
touched the surface on the difficulties in modeling such systems.  In
this paper, we have not addressed the production or feedback of
magnetic fields, which may definitely alter the fate of the fallback.
We have also not discussed the nuclear yields from the ejecta (a first
paper on this is by Fryer et al.  2007 and we plan future projects
studying this nucleosynthesis).  However, note that the old ``wind''
r-process picture will not work in any system with fallback (stars
more massive than $\sim 12$M$_\odot$).  The matter trajectories just
can't be explained by the wind solution in the cases where fallback
dominates the motion near the neutron star.  We also have not studied
the actual accretion onto the neutron star in detail.  Note that the
final neutron star masses in Table~\ref{tab:fallback} assumed no mass
ejecta.  But as we have found in this study, over half of the fallback
material may be ejected, leading to smaller neutron star masses and a
narrower mass range for these neutron stars.

\acknowledgements
It is a pleasure to thank Patrick Young, Frank
Timmes and Aimee Hungerford for useful conversations on this project.
This project was funded in part under the auspices of the
U.S. Dept. of Energy, and supported by its contract W-7405-ENG-36 to
Los Alamos National Laboratory, and by a NASA grant SWIF03-0047.

{}

\clearpage

\begin{deluxetable}{llcccc}
\tablewidth{0pt} 
\tablecaption{Fallback Results\label{tab:fallback}}
\tablehead{ \colhead{Progenitor\tablenotemark{a}} 
& \colhead{Explosion} 
& \colhead{Fallback} 
& \colhead{Neutron Star\tablenotemark{b}} 
& \colhead{Peak Accretion}
& \colhead{Energy} \\ 
\colhead{Mass (M$_\odot$)} 
& \colhead{Energy (10$^{51}$\,erg)} 
& \colhead{Mass (M$_\odot$)} 
& \colhead{Mass (M$_\odot$)} 
& \colhead{Rate (M$_\odot {\rm s^{-1}}$)}
& \colhead{Released (10$^{51}$\,erg)}
}
\startdata

12 & & & & \\
   & 0.55 & 0.33 & 1.64 & $>$0.02 & 120 \\
   & 0.92 & 0.23 & 1.54 & $>$0.01 & 86 \\
   & 2.4 & 0.030 & 1.34 & $>$0.003 & 11 \\
   & 3.6 & 0.015 & 1.33 & $>$0.001 & 5.6 \\
15 & & & & \\
   & 0.9 & 0.34 & 1.75 & $>$0.05 & 130 \\
   & 1.65 & 0.25 & 1.66 & $>$0.05 & 93 \\
   & 7.0 & 0.11 & 1.52 & $>$0.02 & 41 \\
   & 17 & 0.08 & 1.49 & $>$0.01 & 30 \\
23 & & & & \\
   & 1.25 & 2.2 & 3.9 & $>$0.15 & 820 \\
   & 2.0 & 1.75 & 3.45 & $>$0.1 & 650 \\
   & 2.5 & 0.86 & 2.56 & $>$0.1 & 320 \\
   & 6.6 & 0.02 & 1.72 & $>$0.01 & 7.5 \\

\enddata
\tablenotetext{a}{The 12\,M$_\odot$ progenitor is from Heger et al. (2000).  
The 15 and 23\,M$_\odot$ progenitor is from Young et al. (2008).}
\tablenotetext{b}{This is the baryonic mass.  The actual gravitational
mass could be 10\% lower.}

\end{deluxetable}

\clearpage

\begin{deluxetable}{lccccc}
\tablewidth{0pt}
\tablecaption{2-D Simulations\label{tab:neut}}
\tablehead{ \colhead{Model} 
& \colhead{Accretion} 
& \colhead{Boundary\tablenotemark{a}} 
& \colhead{Angular} 
& \colhead{Neutron Star}
& \colhead{Other} \\ 
\colhead{Name}
& \colhead{Rate (M$_\odot$\,s$^{-1}$)} 
& \colhead{Condition} 
& \colhead{Momentum} 
& \colhead{Luminosity ($10^{51} {\rm erg s^{-1}}$)}
& \colhead{}
}
\startdata

NS3 & $10^{-3}$ & $\rho>10^{10},Y_e<0.3$ & $3\times10^{15}{\rm cm^2 s^{-1}}$ & 0. & \\ 
BH3 & $10^{-3}$ & $r<100$\,km & $3\times10^{15}{\rm cm^2 s^{-1}}$ & 0. & \\ 
NS2 & $10^{-2}$ & $\rho>10^{10},Y_e<0.3$ & $3\times10^{15}{\rm cm^2 s^{-1}}$ & 0. & \\ 
NS2-Rot2 & $10^{-2}$ & $\rho>10^{10},Y_e<0.3$ & $10^{16}{\rm cm^2 s^{-1}}$ & 0. & \\
NS2-Hot1 & $10^{-2}$ & $\rho>10^{10},Y_e<0.3$ & $3\times10^{15}{\rm cm^2 s^{-1}}$ & 18. & \\ 
NS2-Hot2 & $10^{-2}$ & $\rho>10^{10},Y_e<0.3$ & $3\times10^{15}{\rm cm^2 s^{-1}}$ & 36. & \\ 
NS2alpha & $10^{-2}$ & $\rho>10^{10},Y_e<0.3$ & $3\times10^{15}{\rm cm^2 s^{-1}}$ & 0. & $2\times \alpha$\tablenotemark{b} \\ 
NS2bound & $10^{-2}$ & $\rho>10^{11},Y_e<0.1$ & $3\times10^{15}{\rm cm^2 s^{-1}}$ & 0. & $2\times \alpha$\tablenotemark{b} \\ 
BH2 & $10^{-2}$ & $r<100\,km$ & $3\times10^{15}{\rm cm^2 s^{-1}}$ & 0. & \\ 
BH2-Rot0 & $10^{-2}$ & $r<100\,km$ & 0. & 0. & \\ 
BH2-Rot2 & $10^{-2}$ & $r<100\,km$ & $10^{16}{\rm cm^2 s^{-1}}$ & 0. & \\ 
BH2-Rot10 & $10^{-2}$ & $r<100\,km$ & $3\times10^{16}{\rm cm^2 s^{-1}}$ & 0. & \\ 
NS1 & $10^{-1}$ & $\rho>10^{10},Y_e<0.3$ & $3\times10^{15}{\rm cm^2 s^{-1}}$ & 0. & \\ 
NS1hot & $10^{-1}$ & $\rho>10^{10},Y_e<0.3$ & $3\times10^{15}{\rm cm^2 s^{-1}}$ & 36. & \\ 
BH1 & $10^{-1}$ & $r<100\,km$ & $3\times10^{15}{\rm cm^2 s^{-1}}$ & 0. & \\

\enddata
\tablenotetext{a}{We have two distinct boundary conditions, one with a limit 
on the density and electron fraction, the other is an absorbing boundary based 
on radius.  We term the absorbing boundary simulations ``BH'' simulations.}
\tablenotetext{b}{To damp out ringing in shocks, smooth particle hydrodynamics 
uses an artificial viscosity.  For most of our simulations, we use the standard 
values for the bulk and von Neumann-Richtmyer viscosities: 1.5 and 3.0 respectively 
(see Fryer et al. 2006 for a review).  In this simulation, we have multiplied 
both these coefficients by 2.}

\end{deluxetable}

\clearpage

\clearpage

\clearpage

\clearpage

\clearpage

\clearpage

\clearpage

\clearpage

\clearpage

\clearpage

\clearpage

\clearpage

\end{document}